\documentclass[aps,prb,reprint,showpacs,superscriptaddress,nofootinbib,floatfix]{revtex4-1}

\usepackage{color}
\usepackage{amsmath}
\usepackage{amssymb}
\usepackage{graphicx}
\usepackage{bm}

\newcommand{\HN}[1]{{#1}}

\newcommand{\HNrev}[1]{{#1}}

\begin{document}

\title{Applications of the Generalised Langevin Equation:\\ 
towards a realistic description of the baths}

\author{H. Ness}\email{herve.ness@kcl.ac.uk}
\affiliation{Department of Physics, Faculty of Natural and Mathematical Sciences,
King's College London, Strand, London WC2R 2LS, UK}

\author{L. Stella}
\affiliation{Atomistic Simulation Centre, School of Mathematics and Physics, Queen's
University Belfast, University Road, Belfast BT7 1NN, Northern Ireland,
UK}

\author{C.D. Lorenz}
\affiliation{Department of Physics, Faculty of Natural and Mathematical Sciences,
King's College London, Strand, London WC2R 2LS, UK}

\author{L. Kantorovich}
\affiliation{Department of Physics, Faculty of Natural and Mathematical Sciences,
King's College London, Strand, London WC2R 2LS, UK}

\begin{abstract}
The Generalised Langevin Equation (GLE) method, {as developed in} 
Ref. [Phys. Rev. B {\bf 89}, 134303 (2014)], 
is used to \HN{calculate} the dissipative dynamics of systems described at the atomic level.
The GLE scheme goes beyond the commonly used bilinear coupling between the 
central system and the bath, and permits us to have a realistic description
of both the dissipative central system and its surrounding bath.
We show how to obtain the vibrational properties of a realistic bath and
how to convey such properties into an extended Langevin dynamics by the use of
the mapping of the bath vibrational properties onto a set of auxiliary variables.
Our calculations for a model of a Lennard-Jones solid show that our GLE scheme 
provides a stable dynamics, with the dissipative/relaxation processes 
properly described.
The total kinetic energy of the central system always thermalises toward the 
expected bath temperature, with appropriate fluctuation around the mean value.
More importantly, we obtain a velocity distribution for the individual atoms in
the central system which follows the expected canonical distribution at the
corresponding temperature. This confirms that both our GLE scheme and our mapping
procedure onto an extended Langevin dynamics provide the correct thermostat.
We also examined the velocity autocorrelation functions and compare our results
with more conventional Langevin dynamics.
\end{abstract}

\pacs{05.10.Gg, 05.70.Ln, 02.70.-c, 63.70.+h}

\maketitle

\section{Introduction}
\label{sec:intro}

Being able to describe the dynamics and dissipation of atomic systems, \HN{modelled}
at the nanoscale, as correctly as possible
is central for modern nanoscience. Nanoscale devices and materials are becoming 
increasingly important in the development of novel technologies. In most applications of these 
new nanotechnologies, the central system is part of a more complex set up where driving forces
are present to establish heat and/or particle flows. The understanding of these corresponding
nonequilibrium properties is of utmost importance. 
This is  especially true when one considers potential applications based on the thermal 
conductivity of materials \cite{Berber2000,Kim2001,Shi2002,Padgett2004,Hu2008,%
Padgett2006,Yang2008,Estreicher2009}
and the heat transport within nanodevices 
\cite{Segal2002,Mingo:2003,Yao:2005,Wang:2007,%
Dubi2011,Widawsky:2012,Cahill2002,Pop2010,Zebarjadi2012}.
Other applications include the study of energy dissipation in solids, or at the
interface between gas phase and a solid phase, and more generally in nanotribology.

In all the examples given above, one has to consider the central open system
surrounded by a heat bath (an environment) which is in contact with the system
and is kept at a given temperature. The general technique
that is specifically appropriate for treating this kind of set up is based
on the so-called Generalised Langevin Equation (GLE)\cite{Mori:1965,Adelman:1976,Adelman:1980,%
Ermak:1980,Carmeli:1983,%
Cortes:1985,Tsekov:1994a,Tsekov:1994b,Risken:1996,Hernandez:1999,Zwanzig:2001,%
Segal:2003,Kupferman:2004,Bao:2004,Izvekov:2006,Snook:2007,%
Kantorovich:2008,Ceriotti:2009,Siegle:2010,Kawai:2011,%
Pagel:2013,Leimkuhler:2013,Baczewski:2013}.
The GLE is an equation of motion for the
non-Markovian stochastic process where the particle (point particle with
mass) has a memory effect to its velocity.

In the conventional Langevin equation, a particle is subjected to a viscous 
drag from the surrounding medium, characterised by some friction force, 
and to a stochastic force that arises because of the coupling of the particle to its surrounding.
The friction constant determines how quickly the system exchanges energy 
with the environment. 
For a realistic description of the surrounding, it is difficult to choose an universal value of the 
friction constant. Indeed each of the vibrational modes in the system would require a different value 
of the friction to be sampled with optimal efficiency. 
Hence a generalization of the conventional Langevin equation is needed, thus leading to the so-called
GLE.

Whenever we are interested in computing properties of materials
at constant temperatures using classical molecular dynamics, it is possible to introduce 
the so-called thermostats, that introduce fluctuations in the total energy consistent with 
the canonical Gibbs sampling of the trajectories of the atoms of the considered
system.
The non-Markovian GLE represents a remarkably flexible framework which permits one to achieve 
a better control over the sampling properties of a molecular dynamics trajectory, to enhance 
its sampling efficiency for all the relevant time scales \cite{Ceriotti:2009,Ceriotti:2010,Morrone:2011,Ceriotti:2011},
to control in a precise manner the disturbance of the dynamics for different frequency ranges
and to provide physical non-equilibrium trajectories for the study of non-equilibrium and/or
relaxation processes.

The GLE has been derived, by one of us, for a realistic system of $N$ particles 
coupled with a realistic (harmonic) bath, i.e. a bath described at the atomic 
level \cite{Kantorovich:2008}.
Non-Markovian dynamics is obtained for the central system with Gaussian 
distributed random forces and a memory kernel 
that is exactly proportional to the 
random force autocorrelation function \cite{Kantorovich:2008}.

Solving the GLE for complex heterogeneous and extended systems is still a challenge,
\HNrev{even when it is known that the GLE dynamics is fully consistent in the sense
that it fulfils the Chapmann-Kolmogorov equations \cite{Gillespie:1996}.} 
A major step towards the solution of this problem for a realistic application has been 
recently given 
in Refs.~[\onlinecite{Ceriotti:2009,Ceriotti:2010,Morrone:2011,Ceriotti:2011,Stella:2014}].
In particular, a very efficient algorithm has been developed in 
Ref.~[\onlinecite{Stella:2014}] to solve the GLE numerically while taking into account 
both fundamental features of the GLE, i.e. a time-dependent memory kernel and the presence 
of a coloured noise which are absolutely essential for a description of the bath at the 
atomic level.

\HN{Such a tool is crucial for the study of nonequilibrium processes in nanoscale systems by 
using molecular dynamics simulations. In the latter, the dissipative processes can be correctly 
described since the system can exchange energy (heat) with the environment. The environment is
characterised by a bath (or several baths), its (their) own dynamical properties going beyond 
conventional thermostats used in classical molecular dynamics (MD) 
simulations \cite{Andersen:1980,Nose:1984a,Nose:1984b,Hoover:1985,Toton:2010}.}

In this manuscript, we present further necessary developments and applications of the method given
in Ref.~[\onlinecite{Stella:2014}].
Specifically, we develop a method and algorithm to calculate the non-Markovian memory
kernel and to perform the mapping of such a kernel onto an extended Langevin dynamics which
permits us to solve the GLE for realistic systems.

The present paper is a proof of principle of the general method described in Ref.~[\onlinecite{Stella:2014}].
As a first application of our method, we consider different model systems that are all
based on a crystalline solid.
For numerical convenience, we model the solid using pairwise Lennard-Jones potentials. 
The calculations should be considered as a robust test of the GLE and methodology
rather than a purely realistic application.

However, in comparison with other GLE implementations, our method includes a realistic coupling
between the central region and the bath which goes beyond the conventionally used
bi-linear coupling. Hence the extended Langevin dynamics developed in Ref.~[\onlinecite{Stella:2014}]
is described with a Verlet-like algorithm which 
{takes full account of}
a functional of the atomic
positions of the central system (which characterises the coupling with the bath). 
The presence of such a functional renders the extended Langevin dynamics equations
highly non-linear in terms of the atomic positions of the central system.

The presented applications are obtained for a ``simple'' model system, but show that our 
scheme is stable and provide the proper description of the essential thermodynamical 
properties of the system, i.e. the proper thermalisation of the system, the proper 
temporal fluctuations of its energy, the proper canonical distributions of the velocities 
and the proper behaviour of the velocity autocorrelation functions.

The paper is organised as follows.
In Sec.~\ref{sec:generalisation}, we recall the central results for the GLE and
how the memory kernel is connected to the polarisation matrix $\bm{\Pi}(\omega)$ which characterises
the vibrational properties of the bath.
Sec.~\ref{sec:PImatrix} is devoted to the scheme we have developed to calculate the
polarisation matrix $\bm{\Pi}(\omega)$ and to map such a central quantity onto a specific analytical
form which permits us to develop an extended Langevin dynamics from the original GLE.
In Sec.~\ref{sec:results}, we provide examples of the calculation and mapping
of the matrix $\bm{\Pi}(\omega)$ for a model of a Lennard-Jones (LJ) solid.
We then use such results to calculate the dynamics of the LJ solid using our extended
phase space GLE dynamics (Sec.~\ref{sec:gle_ex}). We provide results for the 
thermalisation of the system and analyse in detail the corresponding velocity distributions
and velocity autocorrelation functions. We also show how our extended GLE dynamics
is useful in extracting effective friction coefficients for more conventional Langevin
dynamics.
Finally, we discuss further developments and conclude our work in Sec.~\ref{sec:ccl}.

\section{Generalisation and compact form}
\label{sec:generalisation}

\subsection{Heuristic GLE and generalisation}
\label{sec:GLEgeneralisation}

We first start to recall the physical form and contents of the GLE.
For clarity, we consider here
a single degree of freedom (DOF) $q(t)$ with mass $m$ and
momentum $p(t)=m \dot{q}(t)$. 
The corresponding GLE is given by \cite{Mori:1965,Lindenberg:1990}
\begin{equation}
m \ddot{q}(t) = - \partial_q V(q) - \int_{-\infty}^t {\rm d}t' K(t-t') p(t')
+ \eta(t) ,
\label{eq:GLE_1DOF}
\end{equation}
where $V(q)$ is the potential energy, dependent only on the DOF $q$.
The memory Kernel $K(t-t')$ is a characteristic of the bath and the random variable
$\eta(t)$ represents a stochastic process.
The latter is described by a coloured noise and the autocorrelation function of the
stochastic variable is directly
related to the memory kernel, i.e.
$\langle\eta(t)\eta(t')\rangle= k_B T K(t-t')$, where $k_B$ is the Boltzmann constant,
and $T$ the temperature of the system.

{In general, it is difficult to solve the {integro-differential} equation (\ref{eq:GLE_1DOF}),
not only because the atomic momentum needs to be known for all times in the past ($t'<t$),} 
but also because one has to generate a coloured noise $\eta(t)$ that satisfies the 
fluctuation-dissipation relation given above, i.e. the relation linking the noise autocorrelation 
function with the memory kernel.

For some specific analytic forms of the memory kernel, it is possible to
solve exactly the GLE by introducing extra virtual DOF and working with
an extended Langevin dynamics (for all the DOF) involving new stochastic
variables which are then characterised by a white noise distribution \cite{Risken:1996,Zwanzig:2001}.

For example, this can be done with the memory kernel expressed as
a sum of decaying exponentials $K(t-t') = \sum_k e^{-\vert t-t'\vert/\tau_k} c_k/\tau_k$ [\onlinecite{Uhlenbeck:1930}].
Such a Prony series form of the memory kernel has been used to enable an extended variable formalism
in Ref.~[\onlinecite{Baczewski:2013}]. In this case, different characteristic times for relaxation and
dissipation of energy into the bath are used.
A more evolved model can be obtained by taking not only different relaxation processes but also 
some proper internal dynamics of the bath, i.e. the bath is also characterised by some oscillations
of frequency $\omega_k$. In this case the memory kernel has the following form
\begin{equation}
K(t-t') = g^2 \sum_k c^{(k)2}\ e^{-\vert t-t'\vert/\tau_k} \cos(\omega_k\vert t-t'\vert) ,
\label{eq:kernel_model}
\end{equation}
with the constant $g$ representing the strength of the coupling between the system DOF and
the bath.

It can be shown \cite{Ferrario:1979,Marchesoni:1983,Kupferman:2004,Bao:2004,Luczka:2005,Ceriotti:2009} 
that the Generalised Langevin Equation given in Eq.~(\ref{eq:GLE_1DOF}) 
can be conveniently approximated
(for a certain kind of memory kernel) by a Markovian Langevin dynamics 
(with white noise) by introducing a set of auxiliary DOFs.
This approximated equivalence becomes
exact in the limit of infinitely many auxiliary DOFs.
For a memory kernel of
the type given in Eq.~(\ref{eq:kernel_model}), solving the GLE is equivalent to solving 
the following extended variable dynamics \cite{Stella:2014}:
\begin{equation}
\begin{split}
m \ddot{q}(t) & = - \partial_q V(q) + g \sum_k c^{(k)} s^{(k)}_1 \\
\dot{s}^{(k)}_1 & = - \frac{s^{(k)}_1}{\tau_k} + \omega_k s^{(k)}_2 - g c^{(k)} m \dot{q} + \sqrt{\frac{2k_B T m}{\tau_k}} \xi^{(k)}_1 \\
\dot{s}^{(k)}_2 & = - \frac{s^{(k)}_2}{\tau_k} - \omega_k s^{(k)}_1 + \sqrt{\frac{2k_B T m}{\tau_k}} \xi^{(k)}_2 ,
\end{split}
\label{eq:GLE_1DOF_vDOF}
\end{equation}
where the set ${s}^{(k)}_u$ are auxiliary DOF (virtual DOF - vDOF, with ${u=1,2}$) and 
now the stochastic variables are of the white noise type
\begin{equation}
\langle \xi^{(k)}_u(t) \xi^{(k')}_v(t')\rangle = \delta_{uv} \delta_{kk'} \delta(t-t') .
\end{equation}

\HNrev{
In Ref.~[\onlinecite{Stella:2014}], we show how to solve Eq.~(\ref{eq:GLE_1DOF_vDOF})
with the white noise by using a Fokker-Planck (FP) approach. 
The problem is solved in a multivariate form \cite{Gillespie:1996}
and the corresponding probability density 
function is explicitly dependent on the position $q$, momentum $m\dot{q}$ and auxiliary
DOF $s^{(k)}_{u}$. A splitting approach for the corresponding FP propagator is then
used to obtain a (velocity) Verlet-like algorithm to solve the problem.
The dissipative dynamics hence obtained is strictly equivalent to the GLE.
}

A rigorous derivation of the GLE for a complex system made of $N$ atoms (with positions $r_{i\alpha}$
for atom $i$ and Cartesian coordinate $\alpha=x,y$, or $z$)
coupled to a realistic bath has been given by one of us in Ref.~[\onlinecite{Kantorovich:2008}].
Under rather general assumptions concerning the classical Hamiltonian of the system, Equation
(\ref{eq:GLE_1DOF}) can be generalised to many DOF {to mimic the bath}.
Two important assumptions are used in Ref.~[\onlinecite{Kantorovich:2008}]: the fluctuations of the
bath atom positions $u_{l\gamma}$ (for bath atom $l$ Cartesian coordinate $\gamma$) 
are taken to be harmonic around their equilibrium values, and the coupling between
the system and the bath is linear in the bath coordinates. 
The corresponding Lagrangian for the interaction between the system and bath regions is given by
\begin{equation}
\mathcal{L}_{\rm int}(\mathbf{r},\mathbf{u})=-\sum_{l\gamma} \mu_l f_{l\gamma}(\{r_{i\alpha}(t)\})\ u_{l\gamma}(t)
\end{equation}
with $\mu_l$ being the mass of the bath atom $l$. 
Hence, for an arbitrary configuration of the atoms within the
system, there is a force $F_{l\gamma}=\mu_l f_{l\gamma}(\{r_{i\alpha}(t)\})$ acting, at time $t$, 
on the bath DOF $l\gamma$ due to the system-bath coupling.

Under these assumptions, Eq.~(\ref{eq:GLE_1DOF}) is generalised for each $r_{i\alpha}$ and one obtains
a general kernel $K_{i\alpha,i'\alpha'}(t,t')$ which is still related to the noise autocorrelation
function as 
\begin{equation}
\langle\eta_{i\alpha}(t)\eta_{i'\alpha'}(t')\rangle= k_B T K_{i\alpha,i'\alpha'}(t,t'),
\end{equation}
for each noise proces $\eta_{i\alpha}$ associated with the DOF $r_{i\alpha}$.
One should note that now the memory kernel has a full $(t,t')$ time dependence, and not a dependence
on the time difference $t-t'$. This is due to the fact that the system is coupled to the bath via
the function  $f_{l\gamma}(\mathbf{r}(t))$ which is implicitly dependent on time. 

The memory kernel is expressed in the following manner
\begin{equation}
\begin{split}
& K_{i\alpha,i'\alpha'}(t,t') = \\
& \sum_{b,b'}\ {g}_{i\alpha,b}\left(\mathbf{r}(t)\right) \sqrt{\mu_l} \ 
\Pi_{b,b'}(t-t') \sqrt{\mu_{l'}}\ {g}_{i'\alpha',b'}\left(\mathbf{r}(t')\right) ,
\end{split}
\label{eq:genKernel}
\end{equation}
where the quantity $\Pi_{b,b'}(t-t')$ represents the full dynamics of the bath 
{(with indices $b=l\gamma$ and $b'=l'\gamma'$)}.
The quantities ${g}_{i\alpha,b}(\mathbf{r}(t))$ are {obtained} the forces $f_b(\mathbf{r}(t))$
such as ${g}_{i\alpha,b}=\partial f_b(\mathbf{r})/\partial r_{i\alpha}$.

Interestingly, the matrix $\boldsymbol{\Pi}(t-t')$ follows the time translation invariance.
If this matrix could be mapped onto an analytical form of the type given in Eq.~(\ref{eq:kernel_model}),
one could develop a corresponding extended Langevin dynamics for the full GLE.
Such a mapping has been done and derived rigorously in Ref.~[\onlinecite{Stella:2014}] by
using
\begin{equation}
\begin{split}
\Pi_{b,b'}(t-t') \rightarrow \sum_{k=1}^{N_{\rm vDOF}} c_{b}^{(k)} c_{b'}^{(k)}
e^{-\vert t-t'\vert/\tau_k} \cos(\omega_k\vert t-t'\vert) ,
\end{split}
\label{eq:GLE_mapping_vDOF}
\end{equation}
and introducing an extra set of $N_{\rm vDOF}$ auxilliary DOF $s^{(k)}_u$ to solve the GLE in an extended phase space.

Now a few comments are in order. 
On the one hand,
it was shown in Ref.~[\onlinecite{Kantorovich:2008}] that the matrix $\boldsymbol{\Pi}(t-t')$ is 
related to the dynamical matrix of the bath. The solution of the eigenvalue problem for
the dynamical matrix generates the eigenmodes of vibration of the system, with frequency
$\omega_q$ and a corresponding time dependence in  $\cos(\omega_q \vert t-t'\vert)$. Such a result
partially justifies the mapping of $\boldsymbol{\Pi}(t-t')$ as given in Eq.~(\ref{eq:GLE_mapping_vDOF})
as far as the oscillatory behaviour in time is concerned.
Note that the mapping in Eq.~(\ref{eq:GLE_mapping_vDOF}) is used to transform the Langevin dynamics
into an extended phase space where the solution of such a dynamics is more readily accessible.
The mapping in Eq.~(\ref{eq:GLE_mapping_vDOF}) does not necessarily imply that all the 
$\omega_k$ parameters associated with the virtual DOF are all equal to the eigenvalues $\omega_q$ 
of the vibrational modes of the infinite bath region.
Crudely speaking, we can consider the $\omega_k$ as being the frequencies of ``collective'' or 
``coarse-grained'' excitations of the bath. These excitations reduce to the normal modes of the 
bath when one considers as many vDOF as there are DOF in the (actual infinite) bath.

On the other hand, a perturbation introduced in an isolated, finite size, harmonic system cannot
dissipate and the corresponding induced oscillations will survive for ever. However for an
infinite system in the thermodynamic limit, such perturbation will fade away in the long time
limit as the system will equilibrate and return to its thermal equilibrium. In reality,
such a dampening is due to anharmonic effects (phonon-phonon interaction). 
Therefore, the exponential decay of the
$\boldsymbol{\Pi}(t-t')$ matrix is entirely justified in the thermodynamic limit. 
Note that the relaxation times $\tau_k$ are not directly related to the eigenvalues
$\omega_k$ (e.g. like $\omega_k\propto 1/\tau_k$) since they correspond to completely 
different physical processes.

\subsection{Compact matrix form of the GLE}
\label{sec:GLEmatrix}

Using the notation of Ref.~[\onlinecite{Stella:2014}] and the mapping given by Eq.~(\ref{eq:GLE_mapping_vDOF}), 
one can generalise the extended Langevin dynamics for one DOF given by Eq.~(\ref{eq:GLE_1DOF}) to the case
of several DOF in the central system.
In a compact matrix form, the corresponding extended Langevin dynamics is given by
\begin{equation}
\begin{split}
& \mathbf{M}  \ddot{\mathbf{r}} = - \mathbf{\nabla}_\mathbf{r} \bar{V}(\mathbf{r}) + \bar{\mathbf{m}}_B \mathbf{g}(\mathbf{r})\ \mathbf{c}\ \mathbf{s}_1 \\
& \dot{\mathbf{s}}_1 = - \boldsymbol{\tau}^{-1}\mathbf{s}_1 + \boldsymbol{\omega}\mathbf{s}_2 - \mathbf{m}_B \mathbf{g}(\mathbf{r})\ \mathbf{c}\ \dot{\mathbf{r}}
+ \sqrt{2k_BT \bar\mu}\ \boldsymbol{\tau}^{-\frac{1}{2}} \boldsymbol{\xi}_1 \\
& \dot{\mathbf{s}}_2 = - \boldsymbol{\tau}^{-1}\mathbf{s}_2 - \boldsymbol{\omega}\mathbf{s}_1 + \sqrt{2k_BT \bar\mu}\ \boldsymbol{\tau}^{-\frac{1}{2}} \boldsymbol{\xi}_2 ,
\end{split}
\label{eq:GLEmatrix_form}
\end{equation}
where we recall that
$\mathbf{r}$ is a vector of components $r_{i\alpha}$ for all DOF of the system (atom $i$, Cartesian coordinate $\alpha=x,y,z$),
$\mathbf{s}_u$ are vectors with components $s^{(k)}_u$ corresponding to the extra virtual DOF for the extended
Langevin dynamics, with corresponding stochastic vectors $\boldsymbol{\xi}_u$. 
Their components $\xi^{(k)}_u$ obey the Gaussian (white noise) correlation relation: 
\begin{equation}
\langle \xi^{(k)}_u(t) \xi^{(k')}_v(t') \rangle = \delta_{uv} \delta_{kk'} \delta(t-t'), \hspace{5mm}
\langle \xi^{(k)}_u(t) \rangle = 0.
\end{equation}

The quantities $\mathbf{M}, \bar{\mathbf{m}}_B, \mathbf{m}_B$ are diagonal mass matrices with 
elements $m_i \delta_{ij}$ (for the system atom $i$),
$\delta_{ll'} \sqrt{\mu_l/\bar\mu}$ and  $\delta_{ll'} \sqrt{\mu_l\bar\mu}$ respectively, 
where $\bar\mu$ is an effective mass associated with the virtual DOF $s^{(k)}_u$.
The matrix $\boldsymbol{\tau}$ is diagonal, with relaxation time elements $\tau_k$ associated with each vDOF $k$.
 
The potential energy $\bar{V}$ is given by the nominal potential energy $V$ inside the system and 
the potential energy between the system region and the frozen bath region.
There is also a ``polaronic'' correction energy due to the coupling between the system atoms and the harmonic displacements 
of the bath atoms around their equilibrium positions:
\begin{equation}
\begin{split}
\bar{V}(\mathbf{r}) & = V(\mathbf{r}) 
-\frac{1}{2}\sum_{bb'}\sqrt{\mu_l\mu_{l'}}
f_b(\mathbf{r}) \Pi_{bb'}(0) f_{b'}(\mathbf{r}) \\ 
& = V(\mathbf{r}) -\frac{1}{2}
\mathbf{f}(\mathbf{r}) {\mathbf{M}_B}^{-\frac{1}{2}} \boldsymbol{\Pi}(0) {\mathbf{M}_B}^{-\frac{1}{2}} \mathbf{f}(\mathbf{r}),
\end{split}
\label{eq:effec_pot_matrix}
\end{equation}
where we use the indices $b,b'$ for the bath DOF ($b=l\gamma$ for bath atom $l$ and Cartesian coordinate $\gamma$),
and $\mathbf{M}_B$ is a diagonal matrix of the masses of the bath atoms $\mu_l$.
The matrix $\boldsymbol{\Pi}(t-t')$ contains all the information about the dynamics of the bath region and is
related to dynamical matrix of the bath itself. We provide more detail about $\boldsymbol{\Pi}$ in the following
section.

The coupling matrix $\mathbf{g}(\mathbf{r})$ with matrix elements ${g}_{i\alpha,b}(\mathbf{r})$ can be interpreted
as a dynamical matrix between the DOF of the system and the DOF of the bath. 
As mentioned in the previous section, these
matrix elements are obtained from the derivative of the forces acting on the bath DOF with respect to the 
position of the system DOF, i.e. ${g}_{i\alpha,b}=\partial_{i\alpha}f_b(\mathbf{r})$.

Note that, in our notations, the memory kernel $K_{i\alpha,i'\alpha'}(t,t';\mathbf{r})$ entering 
the definition of the GLE is given by
\begin{equation}
\begin{split}
\mathbf{K}(t,t';\mathbf{r}) = \mathbf{g}\left(\mathbf{r}(t)\right) {\mathbf{M}_B}^{\frac{1}{2}} \boldsymbol{\Pi}(t-t') 
{\mathbf{M}_B}^{\frac{1}{2}} \mathbf{g}\left(\mathbf{r}(t')\right) .
\end{split}
\label{eq:kernel_matrix}
\end{equation}

Finally, the properties of the bath are characterised by the matrices $\boldsymbol{\tau}$, $\boldsymbol{\omega}$ 
and $\mathbf{c}$.
They are related to the mapping performed on $\boldsymbol{\Pi}$, see Eq.~(\ref{eq:GLE_mapping_vDOF}), 
to get the extended Langevin dynamics, introduced to solve the GLE.
Since the $\boldsymbol{\Pi}$ depends only on the time difference $\tau=t-t'$, it can be Fourier transformed. 
The mapping of 
$\boldsymbol{\Pi}(\omega)$ is then performed using the following generalised expression \cite{Stella:2014}

\begin{equation}
\begin{split}
\Pi_{b,b'}(\omega)=\sum_{k} c_{b}^{(k)}c_{b'}^{(k)}\left[\frac{\tau_{k}}{1+(\omega-\omega_{k})^{2}\tau_{k}^{2}}\right.\\
\left. +\frac{\tau_{k}}{1+(\omega+\omega_{k})^{2}\tau_{k}^{2}}\right] .
\end{split}
\label{eq:mapping_PI_matrix}
\end{equation}
which is the Fourier transform of $\Pi_{l\gamma,l^{\prime}\gamma^{\prime}}(\tau)$.

\HNrev{
Once more the GLE is solved by considering a multivariate FP problem. The 
corresponding probability density function is now dependent on all 
positions ${\mathbf{r}}$, momenta $\mathbf{M}\dot{\mathbf{r}}$ and auxiliary 
DOFs ${\mathbf{s}}_1$ and ${\mathbf{s}}_2$ [\onlinecite{Stella:2014}].
By using different splitting for the FP propagator, we obtain \cite{Stella:2014}
the algorithm detailed in Appendix \ref{app:algo}.
}

\section{Calculations of the matrix $\mathbf{\Pi(\omega)}$ }
\label{sec:PImatrix}

As shown in Appendix \ref{app:Phproga}, the matrix $\bf\Pi(\omega)$ is related
to the phonon bath propagator $\bm{\mathcal{D}}(\omega)$ as follows:
\begin{equation}
\label{eq:PI_matrix}
\begin{split}
\Pi_{b,b^\prime}(\omega) = - \frac{2}{\vert\omega\vert} {\rm Im} \mathcal{D}_{b,b^\prime}(\omega) ,
\end{split}
\end{equation}
where
the propagator $\bm{\mathcal{D}}(\omega)$ is obtained from the dynamical matrix of the bath $\bm{D}$
as
\begin{equation}
\label{eq:Propaga}
\begin{split}
\mathcal{D}_{b,b^\prime}(\omega) = \left[ \omega^2 \bm{1 - D} + i\varepsilon \right]^{-1}_{b,b^\prime} ,
\end{split}
\end{equation}
with $\varepsilon\rightarrow 0^+$.

The aim of the paper is to develop a robust and efficient numerical scheme to calculate the
inverse of the matrix $\left[ \omega^2 \bm{1 - D} + i\varepsilon \right]$ for an
infinite bath region, or at least for a very large bath region.
It is clear that direct inversion or diagonalisation of the matrix will be very time/resource consuming.

Furthermore, since the bath region will not generally be a fully three-dimensional periodic system, 
a reciprocal $k$-space approach is not necessarily best suited for the problem at hand. 
Hence, we have chosen a more physically intuitive real-space approach based on tridiagonalisation scheme for
inverting the matrix $[ \omega^2 \bm{1 - D} ]$.

\subsection{Real space tridiagonalisation approach}
\label{sec:Lanczos}

We use the Lanczos algorithm,
\begin{equation}
\label{eq:Lanczos}
\begin{split}
\bm{x}_{n+1} = \bm{D} \bm{x}_n - a_n \bm{x}_n - b_n \bm{x}_{n-1} ,
\end{split}
\end{equation}
where the set of coefficients $(a_n,b_n)$ are constructed from the iterative Lanczos
vectors as follows:
 $a_n = \bm{x}^\dag_n\bm{D} \bm{x}_n$ and $b_{n+1} = \vert\vert\bm{x}_{n+1}\vert\vert$ (with $b_0 = 0$,
and before each iteration $\bm{x}_{n+1}$ is renormalised by $1/b_{n+1}$).

The Lanczos algorithm generates the following property:
the $m$-th step of the algorithm transforms the matrix $\bm{D}$ into a tridiagonal matrix
$\bm{T}^{(m)} = \bm{X}^{(m)\dag} \bm{D} \bm{X}^{(m)}$ 
where $\bm{X}_m$ is the transformation matrix whose column vectors are $\bm{x}_0, \bm{x}_1, \bm{x}_2, \cdots, \bm{x}_m$.
The tridiagonal matrix has diagonal elements $[\bm{T}^{(m)}]_{n,n}=a_n$
and off-diagonal elements  $[\bm{T}^{(m)}]_{n+1,n}=[\bm{T}^{(m)}]_{n,n+1}=b_{n+1}$.
It is then easier to calculate the inverse of a matrix when it is given in a tridiagonal form
since it can be expressed as a continued fraction.

In order to obtain the diagonal elements of $[ \omega^2 \bm{1 - D} ]^{-1}_{b,b}$, one starts the Lanczos algorithm
with an initial Lanczos vector $\bm{x}_0 = \bm{u}_b$. 
The vector $\bm{u}_b$ is a unit vector in the corresponding vector-space. 
\HN{The vector has a length of $3\times(N_B+N_{\rm at} )$ where $N_B$ is the number of atoms in the
bath region and $N_{\rm at}$ the number of atoms in the central system.
The vector $\bm{u}_b$  has all elements $u_b[j]=0$ apart from the component $i$ of interest for which $u_b[i]=1$
and} which
corresponds to the $l$-th bath atom with Cartesian coordinate $\gamma$ ($b=l\gamma$).

After tridiagonalisation, we then obtain the element $[ \omega^2 \bm{1 - D} ]^{-1}_{b,b}$ as a continued
fraction:
\begin{equation}
\begin{split}
\bm{x}^\dag_0 [ \omega^2 \bm{1 - D} ]^{-1}\bm{x}_0 & = \\
[ \omega^2 \bm{1 - D} ]^{-1}_{b,b} & = 
\cfrac{1}{\omega^2 - a_0 - \cfrac{b_1^2}{\omega^2 - a_1 - \cfrac{b_2^2}{\omega^2 - a_2 - \dots }   } } 
\end{split}
\label{eq:PIbb_contfrac}
\end{equation}

In order to calculate the off-diagonal elements $[ \omega^2 \bm{1 - D} ]^{-1}_{b,b'}$, one performs
two Lanczos iterations starting with two different initial Lanczos vectors $\bm{x}^\pm_0 = (\bm{u}_b \pm \bm{u}_{b'}) / \sqrt{2}$.
The off-diagonal elements are extracted from the difference of two continued fractions obtained since
\begin{equation}
\begin{split}
& [ \omega^2 \bm{1 - D}  ]^{-1}_{b,b'} = \\
& \frac{1}{2} \left(
\bm{x}^{+\dag}_0 [ \omega^2 \bm{1 - D} ]^{-1} \bm{x}^+_0 - \bm{x}^{-\dag}_0 [ \omega^2 \bm{1 - D} ]^{-1} \bm{x}^-_0 
\right),
\end{split}
\label{eq:PIbbp_contfrac}
\end{equation}
and the dynamical matrix is symmetric $[\bm{D}]_{b,b'}=[\bm{D}]_{b',b}$.

With this procedure we can calculate all matrix elements $\Pi_{b,b'}$ from Eq.~(\ref{eq:PI_matrix}).
Another advantage of using the Lanczos iterative scheme in comparison with exact diagonalisation or inversion comes
from the following fact: the correct results are obtained once the coefficients of the continued fraction have
converged towards an asymptotic value. 
For the system we have considered (see the next section), the convergence is always reached for a level $M$ 
of the continued fraction much smaller than the \HN{dimension of the dynamical matrix 
$N_D = 3\times(N_B+N_{\rm at} )$}.
One of the reasons for that is that the range of the inter-atomic interaction is finite and therefore
the off-diagonal elements of the dynamical matrix decrease with the inter-atomic distance between the bath DOF
$b$ and $b'$ quite rapidly (at least for a non-ionic system).
In terms of scaling, the Lanczos scheme appears more efficient since exact diagonalisation or inversion scales
as $N_D^3$ while the Lanczos iterations involve only matrix-vector multiplication, scaling as $N_D^2$.

\subsection{Mapping the $\Pi_{bb'}(\omega)$ matrix}
\label{sec:mapping}

Once a model atomic configuration for the bath region is chosen, the corresponding dynamical matrix
can be calculated numerically.
Note that in calculating the dynamical matrix of the bath region which is surrounding the central region 
(the system), we have to consider the interactions between the bath atoms and the central region as well. 
In doing so, the atoms in the central system can be placed at their equilibrium positions.

From the knowledge of the dynamical matrix,
we can calculate all the matrix elements $\Pi_{b,b'}$ using the Lanczos scheme and then perform the mapping 
expressed by Eq.~(\ref{eq:mapping_PI_matrix}). 
We perform this mapping by fitting the calculated $\Pi_{b,b'}(\omega)$
functions onto the sums of Lorentzian functions given by Eq.~(\ref{eq:mapping_PI_matrix}).

Once the mapping is performed, the set of parameters $c_b^{(k)}, \omega_k$ and $\tau_k$ characterising the
vibrational properties of the bath region can be used for any extended GLE dynamics of the central system region. 
The calculations outlined here for the virtual DOF associated with the bath are done before performing
any extended GLE dynamics for various systems coupled to this bath and for different bath temperatures.

There are different ways to perform the fit needed for the mapping. 
One could perform a direct brute-force fit
of the $\Pi_{b,b'}(\omega)$ functions ($N_B (N_B+1)/2$ functions) altogether 
onto the analytical expression used for the mapping and extracting the relevant parameters 
$c_b^{(k)}, \omega_k$ and $\tau_k$. 
This is however a highly complex task
as we have found that reaching local minima on generalised trajectories in the corresponding phase-space 
may be impossible to achieve without knowing more about
the location of the expected target in the corresponding phase-space.

The mapping procedure given below is one of many possible approaches, including conjugated
gradient or the Levenberg-Marquardt algorithm for damped least-square minimization which are under
consideration \cite{SummerStudents:2014} or ``compressed sensing'' fitting algorithms \cite{Andrade:2012}.

In this manuscript, we use a different method based on a more intuitive physical approach
which can be summarised as follows.
For a finite size bath, the exact $\Pi_{b,b'}(\omega)$ is given by a series of peaks whose positions/amplitudes
are related to the eigenvalues/vectors of the dynamical matrix. By introducing a Lorentzian broadening of 
these peaks, the mapping shown in Eq.~(\ref{eq:mapping_PI_matrix}) is exact when the $\omega_k$ parameters 
are taken to be the eigenvalues of the dynamical matrix, the $c_b^{(k)}$ parameters are the components 
of the eigenvectors on the basis of the bath DOF $b$, while the $\tau_k$ parameters are related to the width 
of these peaks.
For an infinite system characterising a realistic bath in the thermodynamic limit, one would get an infinite
number of eigenvalues/vectors, and the mapping in Eq.~(\ref{eq:mapping_PI_matrix}) becomes approximate since
we consider only a finite number of virtual DOF. 
In this case, the mapping corresponds to a ``coarse-grained'' description of the bath.

 
Hence we have devised the following fitting procedure of the $\Pi_{b,b'}(\omega)$ functions (examples of
the corresponding mapping are given in the next section).

\begin{itemize}
\item
Find, numerically, the position of all peaks in all diagonal elements $\Pi_{b,b}(\omega)$.
\item
Conserve the most relevant peaks $\omega_k$ and eventually add extra peaks, on a denser
mesh, around $\omega\rightarrow 0$ if necessary. 
This is a user-dependent choice, the only one in the mapping procedure.
It is very important as it determines the number of virtual DOF $N_{\rm vDOF}$.
\item 
For all the diagonal elements  $\Pi_{b,b}(\omega)$, use a least-square fit to determine
the amplitude $A_b^{(k)}=\vert c_b^{(k)}\vert^2$ and width $1/\tau_b^{(k)}$ of each peak 
corresponding to the virtual DOF $k=1,2,\dots,N_{\rm vDOF}$.
\item
From the mapping Eq.~(\ref{eq:mapping_PI_matrix}), the $\tau_k$ is independent of the
bath DOF index, hence take $\tau_k = \text{min}_b \{ \tau_b^{(k)} \}$.
\item
Determine the sign of the coefficient $\pm\vert c_b^{(k)}\vert$ from a best fit
on all the $N_B (N_B-1)/2$ off-diagonal elements $\Pi_{b,b'}(\omega)$.
\end{itemize}

\begin{figure}
\begin{centering}
\includegraphics[width=70mm]{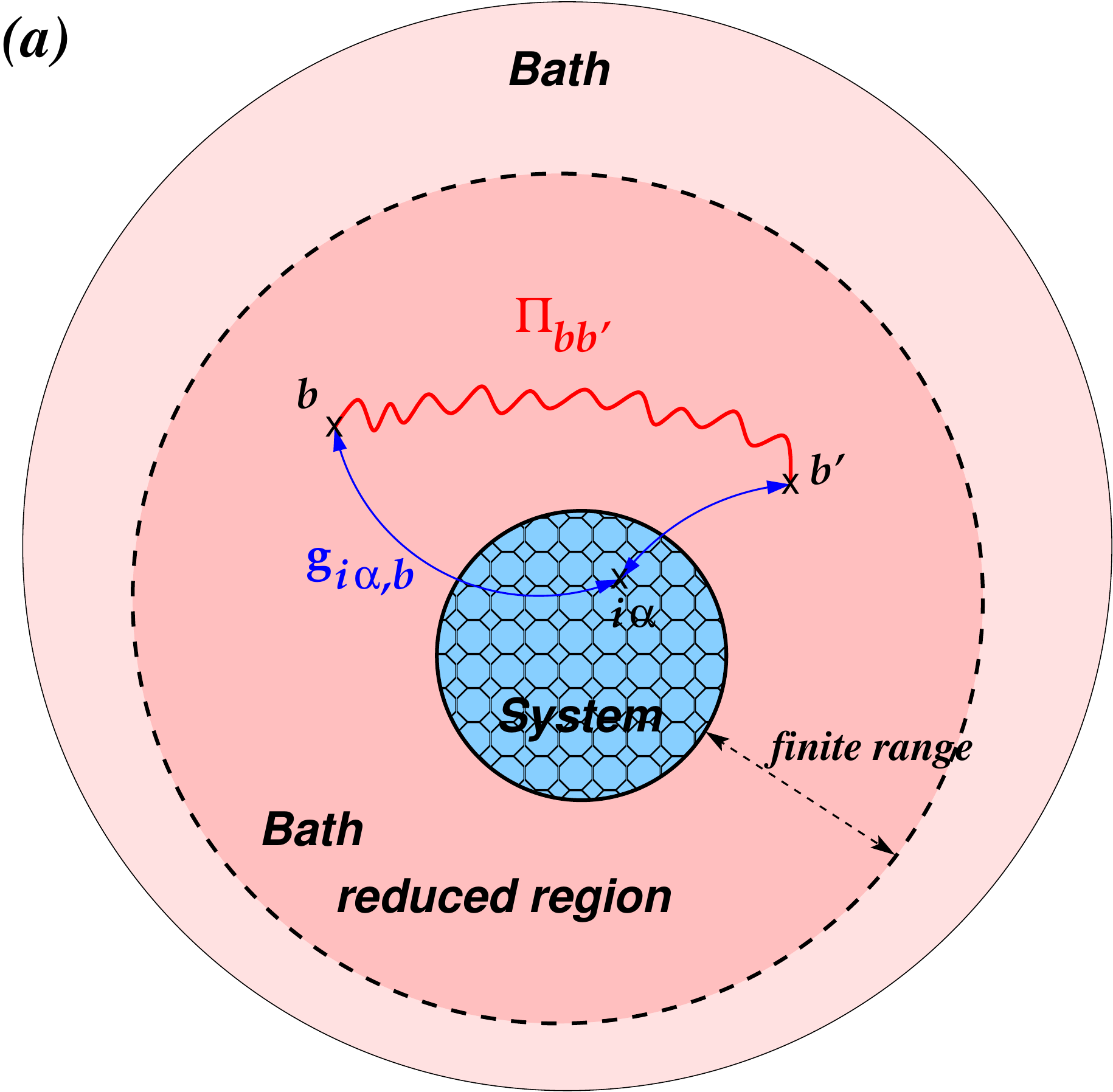}\\
\includegraphics[width=75mm]{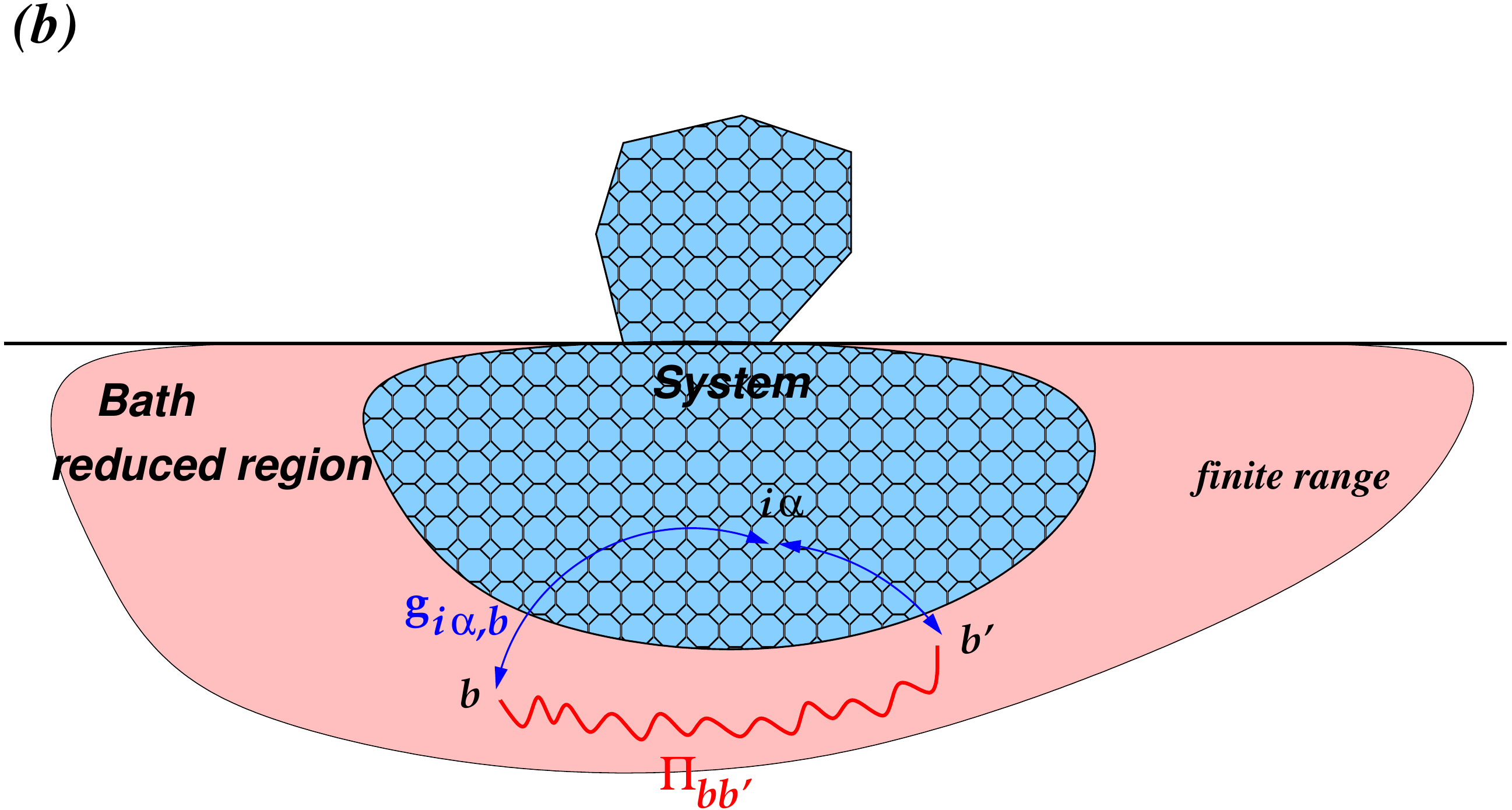}
\end{centering}
\caption{(Colour online) Schematic representation of total systems under consideration. 
This includes the finite size central system (blue)
where the GLE dynamics is performed, and the bath region (pink). 
Because the forces $f_b$ and the quantities $g_{i\alpha,b}$ are of finite range (not
necessarily short ranged), one can perform the mapping of $\Pi_{bb'}(\omega)$ on a finite region
of space (the bath reduced region). 
Furthermore, the matrix elements $\Pi_{bb'}(\omega)$ go to zero when the distances
between the two bath DOF $b$ and $b'$ become large.
\HN{The central system contains $N_{\rm at}$ atoms, the bath region $N_B$ atoms and the reduced
bath region $N_B^{\rm red}$ atoms respectively.} }
\label{fig:systems}
\end{figure}

The algorithm devised above is just one of the many possible ways of performing the mapping.
Our choice clearly emphasizes a better fit for the diagonal elements $\Pi_{b,b}(\omega)$
for the mapping Eq.~(\ref{eq:mapping_PI_matrix}). The choice of determining the
sign $\pm\vert c_b^{(k)}\vert$ is reminiscent of the results obtained for a finite
size system, where the $c_b^{(k)}$ parameters would be equivalent to the components of the 
corresponding eigenvectors of the dynamical matrix.

{Finally, one should note that
since the forces $f_b$ and the quantities $g_{i\alpha,b}$ are of finite range (not
necessarily short ranged), the Kernel built on the quantities  $g_{i\alpha,b}$
and $\Pi_{b,b'}$, see Eqs.~(\ref{eq:genKernel}) and (\ref{eq:kernel_matrix}), 
does not need to be computed by means of infinite sums on the bath indices $b$ and $b'$. 
Therefore we can reduce the number of $\Pi_{b,b'}$ components to be calculated.
We perform the mapping of $\Pi_{b,b'}(\omega)$ on a finite region of space which
we call the bath reduced region as shown in Figure \ref{fig:systems}.
Although this was the strategy adopted in the present study, the bath region used 
for the mapping and the summation in  Eq.~(\ref{eq:genKernel}) with respect to the
bath sites may not necessarily be the same, e.g. one may use a larger bath region
for the mapping to have a better representation for the bath when fitting the 
parameters (and the number) of the vDOF.}

\section{Results for the $\boldsymbol{\Pi}(\omega)$ matrix}
\label{sec:results}

\subsection{Calculation of the polarisation matrix $\bm{\Pi}$ }
\label{sec:calc_PI_mat}

As a first step in the application of our method, we have implemented the
procedure described above in the classical MD code LAMMPS \cite{Plimpton:1995}.
Such a procedure is best suited to study the dissipative dynamics of the systems
schematically depicted in Figure \ref{fig:systems}. These systems are typically
either a bulk-like cluster (containing defects or not) coupled to its three
dimensional surrounding as shown in panel (a) of Fig.~\ref{fig:systems}, or 
any kind of structures deposited on a surface as shown in panel (b).

Once the total system is built with a clear distinction between the central system region 
and the bath region, we calculate the dynamical matrix using numerical differentiation of 
the forces acting on bath atoms obtained from LAMMPS. Note that, as mentioned previously, 
we consider for such calculations the whole system made of the central system and the bath 
region. 
The dynamical matrix is obtained from the conventional expression:
\begin{equation}
\begin{split}
D_{b,b'} = \frac{1}{\sqrt{\mu_l \mu_{l'}}} 
\frac{\partial^2 E_{\rm tot}}{\partial{r_{l\gamma}} \partial{r_{l'\gamma'}}}
= \frac{1}{\sqrt{\mu_l \mu_{l'}}} \frac{\partial f_{b'}}{\partial{r_{l\gamma}}}
\end{split}
\label{eq:dynmat}
\end{equation}
In all our calculations, we have verified that the acoustic sum
rule is fulfilled, i.e. $\sum_b D_{b,b'} = \sum_{b'} D_{b,b'} = 0$.

\begin{figure}
\begin{centering}
\includegraphics[width=40mm]{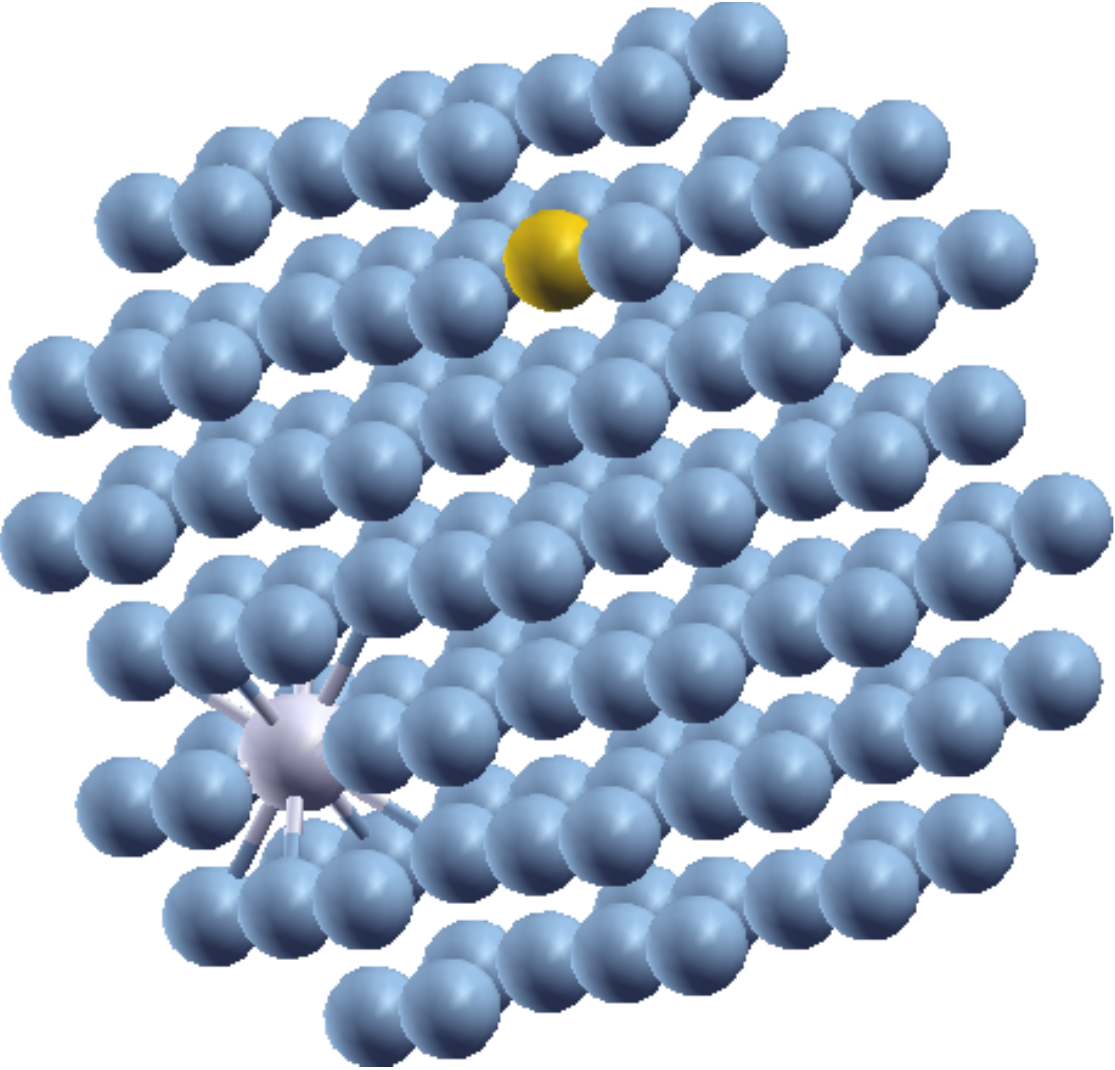}
\includegraphics[width=35mm]{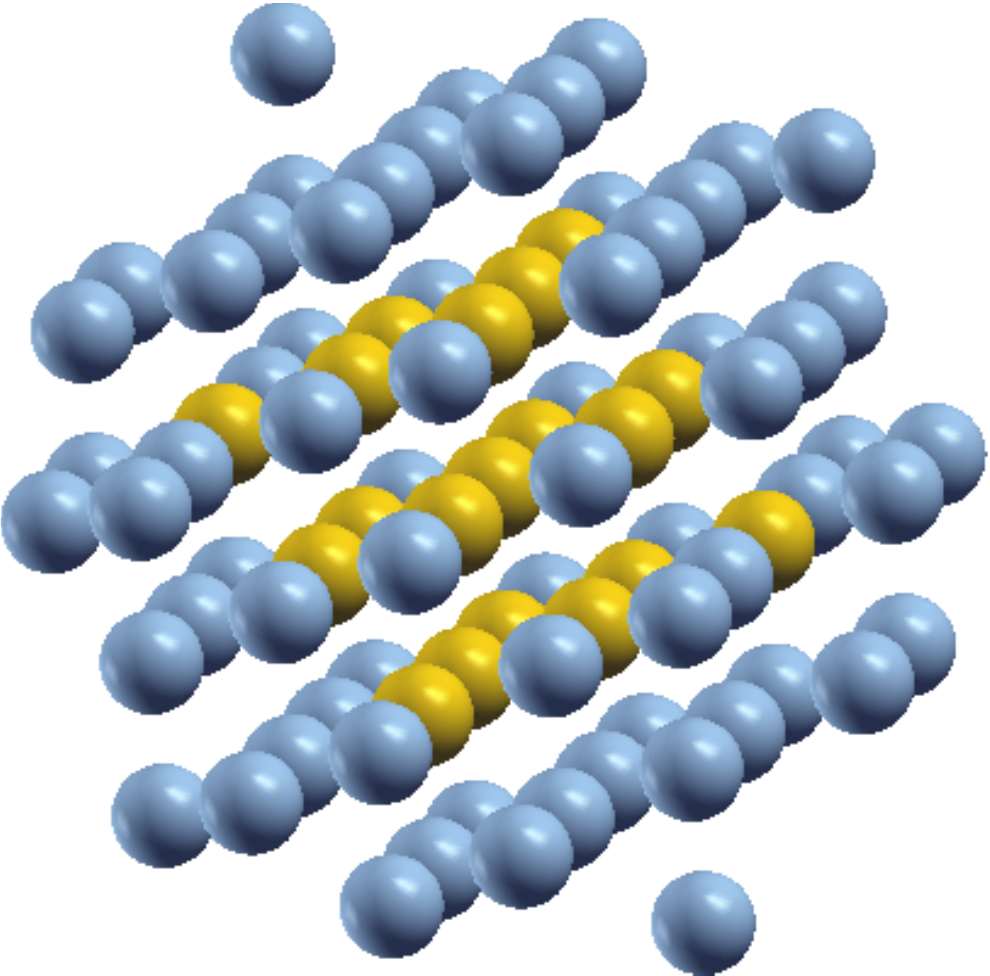}
\end{centering}
\caption{(Colour online) \HN{Model system of a LJ solid, fcc lattice.
(Left) System for the calculation of the dynamical matrix 
and for the mapping. It contains the central system (made of $N_{\rm at}$ atoms) 
and the entire bath region (made of $N_B$ atoms). It has
a corresponding radius of $R = 7.5$ \AA\ and $N_B+N_{\rm at}=135$. 
The bath atom labelled $l$ is coloured in yellow, and the 
bath atom $l'$ is in light-grey (surrounded by interatomic bonds for clarity).
(Right) System for the GLE calculations. It consists of
the system region containing $N_{\rm at}=19$ atoms (yellow), and of the bath
reduced region containing $N_B^{\rm red}=68$ atoms (grey-blue).} }
\label{fig:system1}
\end{figure}

To validate our methodology, we show, in this paper, results for the mapping of the 
$\Pi_{bb'}$ matrix and for the
corresponding GLE dynamics for a simple model of a Lennard-Jones (LJ) solid.
The interaction between every pair of atoms $(i,j)$ at the distance $r_{ij}$ is
given by the conventional LJ potential 
$V(r_{ij}) = 4 \epsilon \left[ (\sigma/r_{ij})^{12} - (\sigma/r_{ij})^6 \right]$.
For convenience, we take the LJ parameters ($\epsilon = 0.583$ eV and $\sigma = 2.77$ \AA) 
for a solid built as a fcc lattice with the lattice parameter
$a_0 = 4.025$ \AA\ (i.e. the nearest neighbour distance $d_{\rm NN}=\sqrt{2}/2 a_0 = 2.85$ \AA)
[\onlinecite{note:1}].
In the following, we show results obtained from the dynamical matrix of the cluster made of 135
atoms (left panel in Fig.~\ref{fig:system1}).

\begin{figure}
\begin{centering}
\includegraphics[width=80mm]{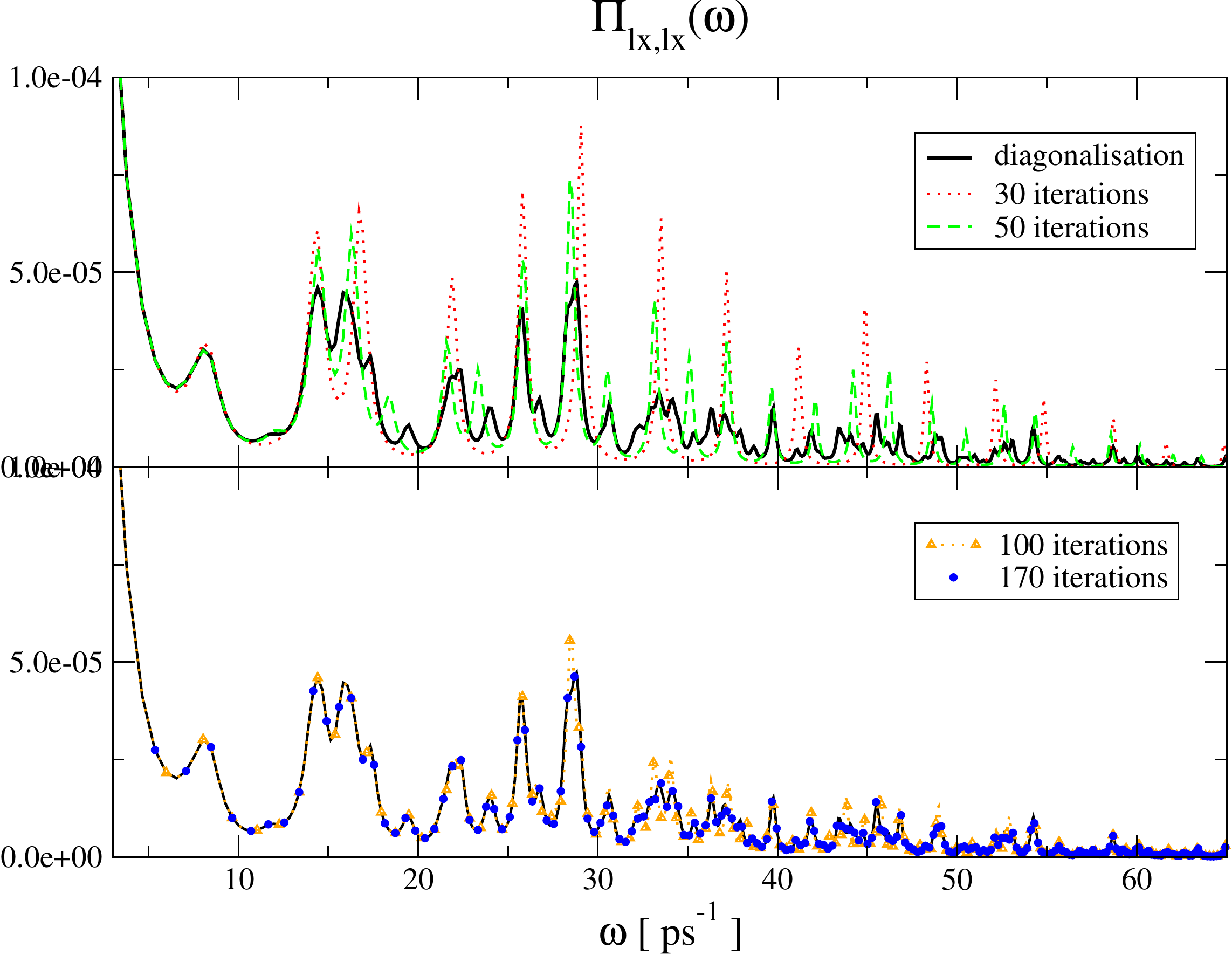}
\end{centering}
\caption{(Colour online) Diagonal matrix element $\Pi_{b,b}(\omega)$ for $b = l x $, 
with atom $l$, shown in the left panel of Fig.~\ref{fig:system1}, calculated from exact 
diagonalisation of the
dynamical matrix, and from the Lanczos iterative scheme using a different number of iterations.
For this example, one gets good results after 100 iterations which is still much smaller 
than $N_{\rm dim}$.
Calculations are performed with a small imaginary part $\varepsilon=3$.
Note that, from the definition given in Eq.~(\ref{eq:Dwitheta}),
the value of $\varepsilon$ has to be compared with the typical $\omega_\lambda^2$ values.
Only the $\omega\ge 0$ part of the functions is shown here and below, 
since $\Pi_{b,b'}(\omega)$ is an even function.}
\label{fig:PIbb_Lanc_iter}
\end{figure}

First we test the convergence of the calculation of $\Pi_{b,b'}$ 
with respect to the number of Lanczos iterations.
Figure \ref{fig:PIbb_Lanc_iter} shows typical results for the diagonal matrix element
$\Pi_{b,b}$ (here $b \equiv l x $ with atom $l$ shown in the left panel of Fig.~\ref{fig:system1}).
As expected, increasing the number of Lanczos iterations allows us to convergence towards
the exact result for $\Pi_{b,b'}$ obtained from direct diagonalisation. What is 
very interesting and useful for numerical applications, is that  $\Pi_{b,b'}$ can be
obtained with a good level of accuracy from a number of iterations much smaller than the
actual dimension $N_{\rm dim}$ of the dynamical matrix. We suspect that such a behaviour
arises from the structure of the dynamical matrix, which presents the form of a sparse
matrix.
This is typical for a system with interaction of a finite range; however, a similar result 
may not hold for a system in which the interaction between atoms is dominated by long-range 
Coulomb interactions.

\begin{figure}
\begin{centering}
\includegraphics[width=80mm]{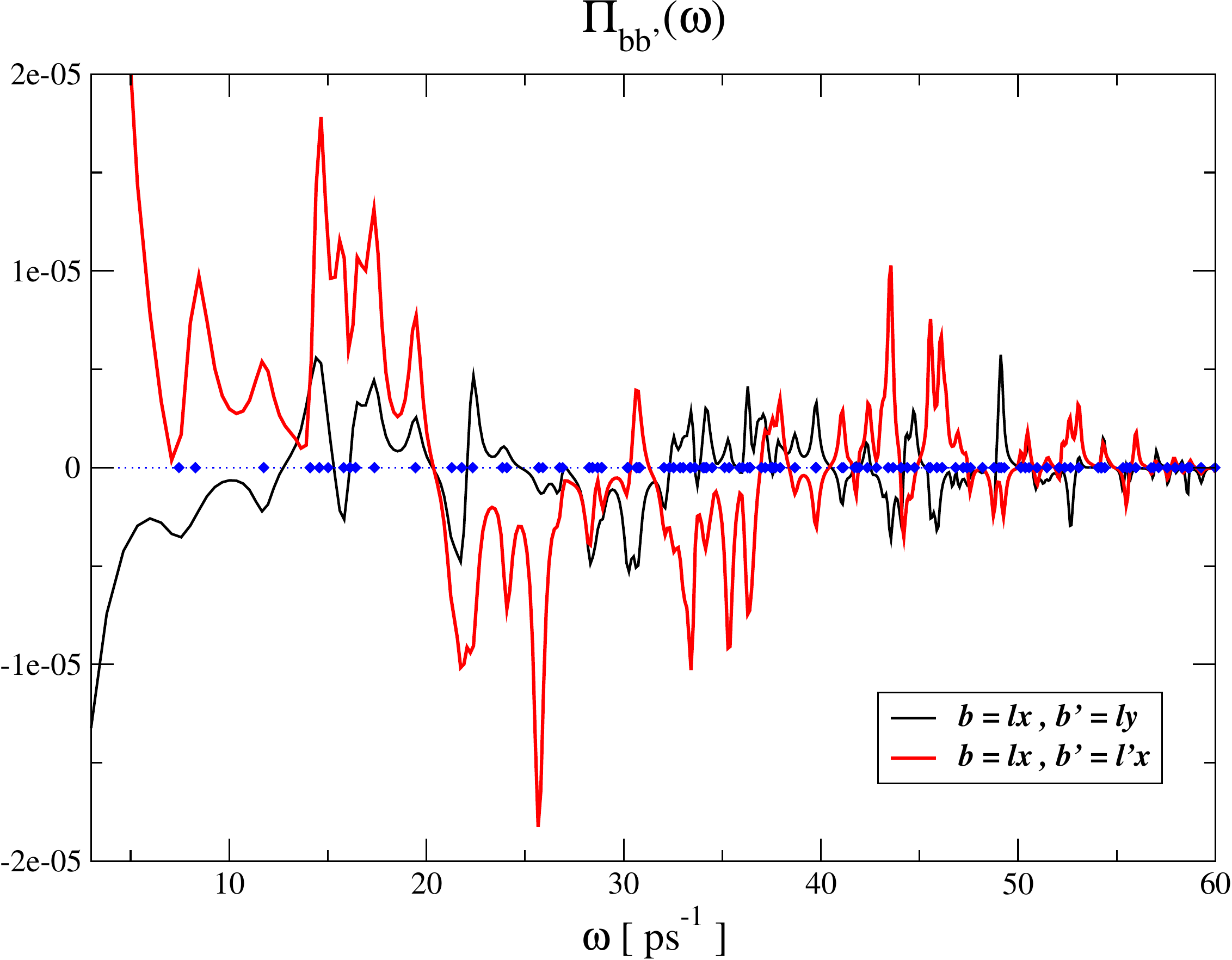}
\end{centering}
\caption{(Colour online) Examples of two off-diagonal matrix elements $\Pi_{b,b'}(\omega)$ corresponding
to a local (in space) matrix element $b = l x$, $b = l y$ , 
and a non-local matrix element $b = l x$, $b = l'x$ where
the distance between the two atoms $l$ and $l'$ is $d_{ll'} = 2.12 a_0 = 8.54$ \AA.
The diamond symbols represent the exact eigenvalues of the corresponding dynamical matrix.
Calculations are performed with a small imaginary part $\varepsilon=3$.
The two atoms corresponding to the DOF $b$ and $b'$ are shown in the left panel
of Fig.~\ref{fig:system1}. }
\label{fig:PIbbp_Lanc_example}
\end{figure}

Figure \ref{fig:PIbbp_Lanc_example} shows some typical examples for the off-diagonal matrix 
elements $\Pi_{b,b'}$ obtained from converged Lanczos iterations.
As expected, the off-diagonal elements have both positive and negative contributions, only the 
diagonal matrix elements are positive functions of $\omega$.
Furthermore, each peak in the $\Pi_{b,b'}(\omega)$ functions 
(as well as for the diagonal $\Pi_{b,b}(\omega)$ functions) corresponds to an eigenvalue of 
the dynamical matrix. 
Note that it does not imply that all eigenvalues are necessarily represented by peaks in 
any $\Pi_{b,b'}(\omega)$ functions.

Another important point concerns the amplitude of the $\Pi_{b,b'}(\omega)$ functions:  
{the amplitude of the off-diagonal elements is much smaller than the amplitude 
of the diagonal ones}
(at least one order 
of magnitude smaller for the examples shown in Fig.~\ref{fig:PIbb_Lanc_iter} and 
Fig.~\ref{fig:PIbbp_Lanc_example}). This is even
more true when the spatial separation $d_{ll'}$ between the two bath DOF $b$ and $b'$ becomes 
larger ($d_{ll'} \gg a_0$).
Such a behaviour justifies {\it a posteriori} the fact that one does not need to consider all the matrix
elements of an infinite bath to be able to describe properly its intrinsic vibrational properties.

\begin{figure}
\begin{centering}
\includegraphics[width=80mm]{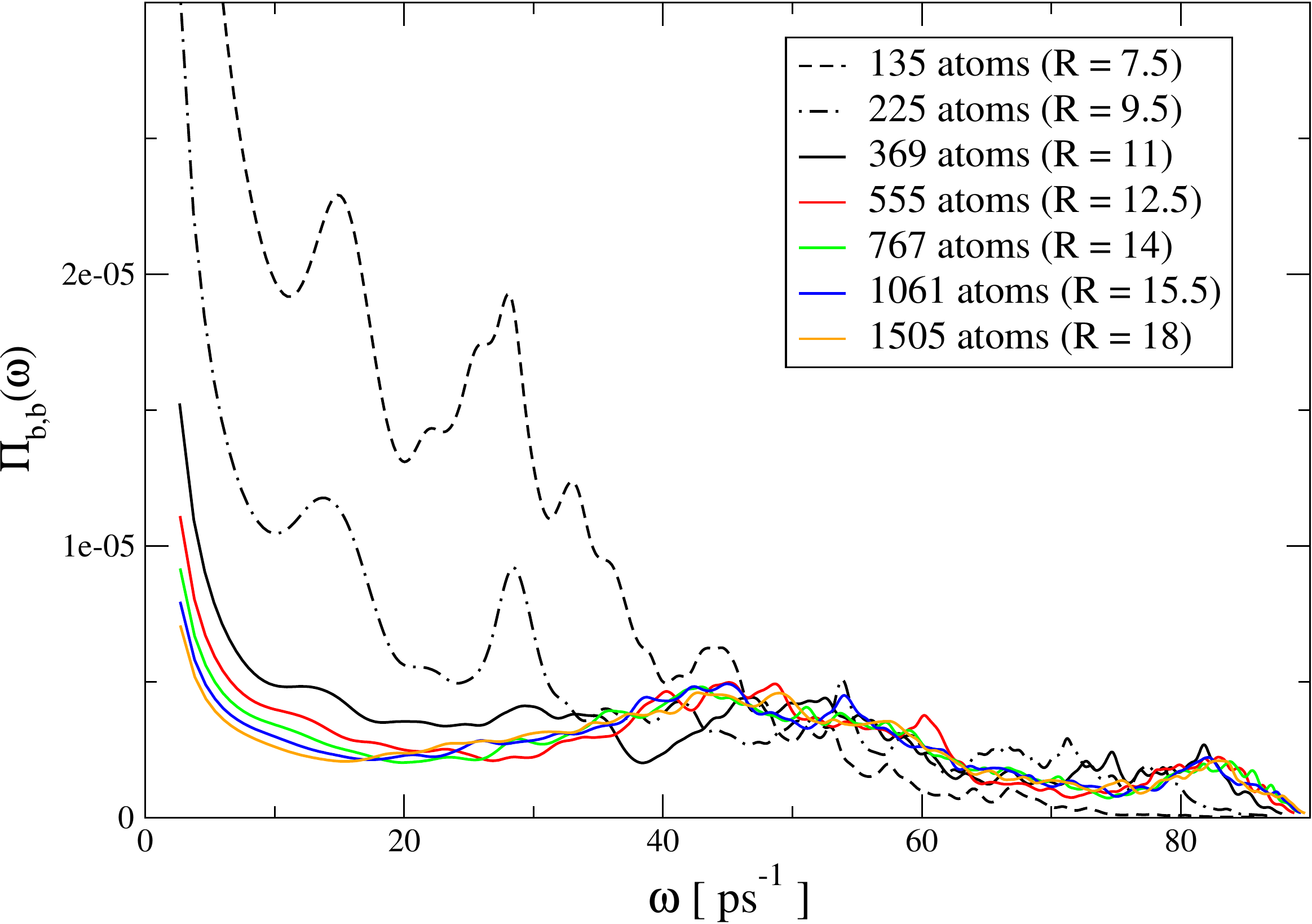}
\end{centering}
\caption{(Colour online) Convergence of the diagonal matrix element $\Pi_{b,b}(\omega)$ 
with respect to the size of the bath region.
{ The bath DOF $b = l x$ is the yellow atom in the cluster made of 135 atoms (radius $R = 7.5$ \AA)
as shown in the left panel of Figure~\ref{fig:system1}}.
By adding extra layers of bath atoms, the size of the cluster increases further
from 225 atoms (radius $R = 9.5$ \AA), 
369 atoms ($R = 11$ \AA), 
555 atoms ($R = 12.5$ \AA), 
767 atoms ($R = 14$ \AA),
1061 atoms ($R = 15.5$ \AA)
to
1505 atoms ($R = 18$ \AA).
The convergence of the lineshape of the matrix element $\Pi_{b,b}(\omega)$
is achieved for a bath region of radius $R \ge 12$ \AA.
Calculations are performed with an imaginary part $\varepsilon=9$ to obtain smooth
curves.}
\label{fig:PIbb_vs_bathsize}
\end{figure}

We now study the convergence properties of the $\Pi_{b,b'}(\omega)$ versus the size of
the considered bath region.
This is important as increasing the size of the cluster considered in the Lanczos
procedure makes more remote atoms of the bath to be available to the Lanczos
iterations.
For that, we consider one $\Pi_{b,b}(\omega)$ for one fixed bath index $b$ located inside 
the bath reduced region (see the yellow atom in the 135 atoms cluster with a radius 
of $R = 7.5$ \AA\ shown in the left panel of Fig.~\ref{fig:system1}).
We then add extra layers of atoms to this cluster to simulate a larger bath region.
The convergence of the $\Pi_{b,b}(\omega)$ function is shown in Figure~\ref{fig:PIbb_vs_bathsize}.
The convergence in the lineshape of the matrix element $\Pi_{b,b}(\omega)$
is achieved for a bath region of radius $R \ge 12$ \AA, which corresponds to $R \sim 3 a_0$.
These results show that the vibrational properties of the bath are more long-ranged
than initially expected. 
We believe that the convergence does depend on the range of the pair-wise
potential, which in our case is modelled with a cut-off of $R_{\rm cutoff} = 6.5 \sim 1.6 a_0$.

Finally, we would like to comment on the behaviour of the $\Pi_{b,b'}(\omega)$ functions in the
limit of $\omega\rightarrow 0$. The lowest frequency behaviour seems to be like $\pm 1/\omega^a$
(with $a\sim 1.0$). 
In principle, one would expect a finite value for $\Pi_{b,b}(\omega\rightarrow 0)$
as was shown analytically in Ref.~[\onlinecite{Stella:2014}] for a simple one-dimensional model.
We argue that the behaviour at small $\omega$ we observe in our numerical simulations
is due to a finite-size effect. The acoustic long-ranged vibrational properties 
of a solid are not appropriately well described using finite-size cluster dynamical matrix calculations.
This is clear from Fig.~\ref{fig:PIbb_vs_bathsize} that such a behaviour becomes less and less 
dominant in the lineshape of $\Pi_{b,b}(\omega)$ function when the size of the system increases.
The larger systems are considered, the better the description of the low-frequency, long wavelength
vibrations will be.

However, we want to stress that such low frequency acoustic modes are not the vibrational modes
which will be dominant in the dissipation processes between the system and the bath regions.
In the following sections, we show that an approximate description of the low frequency range
of the $\Pi_{b,b'}(\omega)$ functions does not lead to the wrong physical behaviour of the dynamics
of the systems obtained from the GLE, at least for not too long MD runs.

\subsection{Fitting the diagonal elements of $\Pi_{b,b}(\omega)$}
\label{sec:fitPIbb}

Once, we have chosen the number of vDOF we want to work with, the fitting procedure
described in Sec.~\ref{sec:mapping} is used to map the diagonal elements
$\Pi_{b,b}(\omega)$ according to the expression given in Eq.~(\ref{eq:mapping_PI_matrix}).

\begin{figure}
\begin{centering}
\includegraphics[width=80mm]{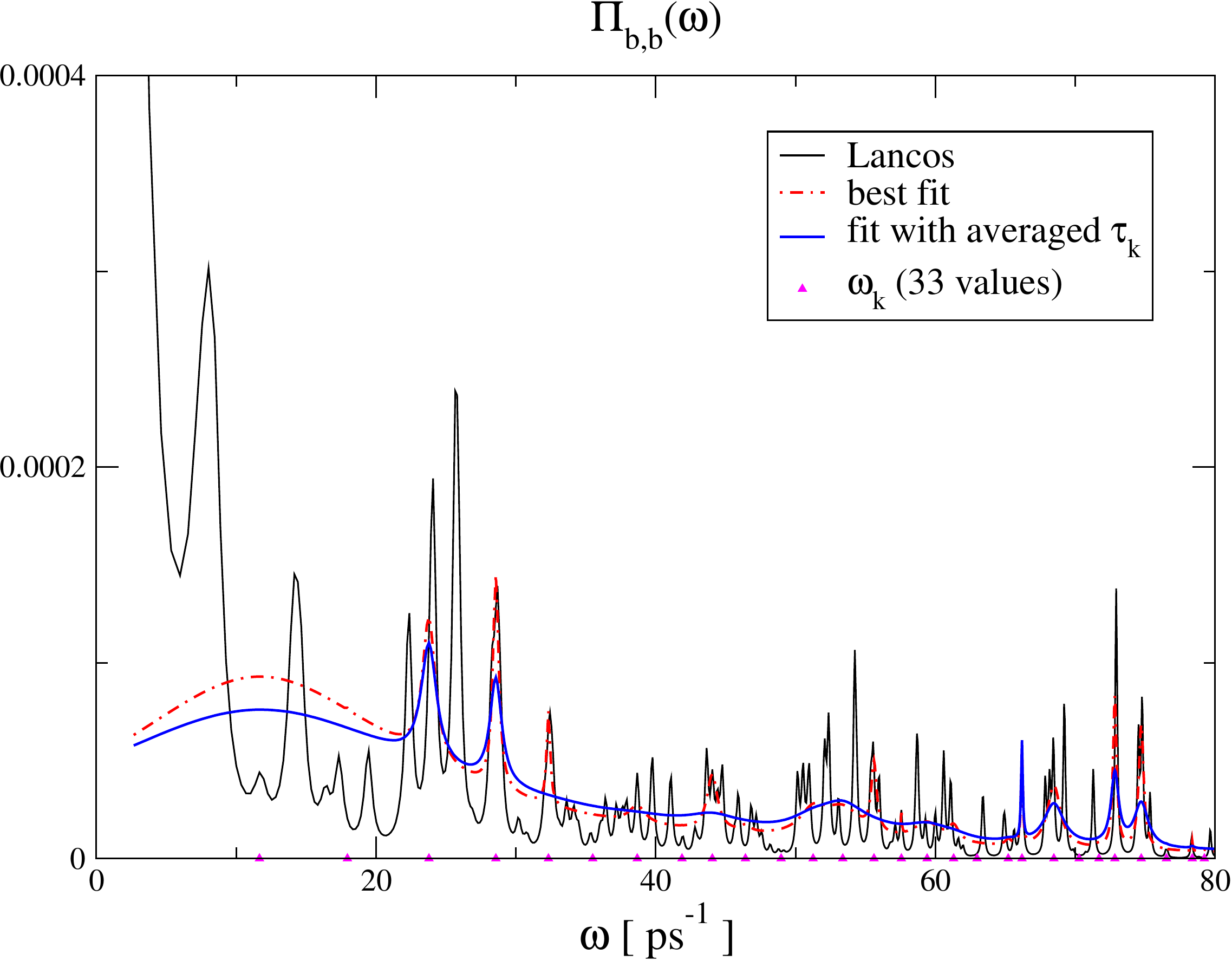}\\
\includegraphics[width=80mm]{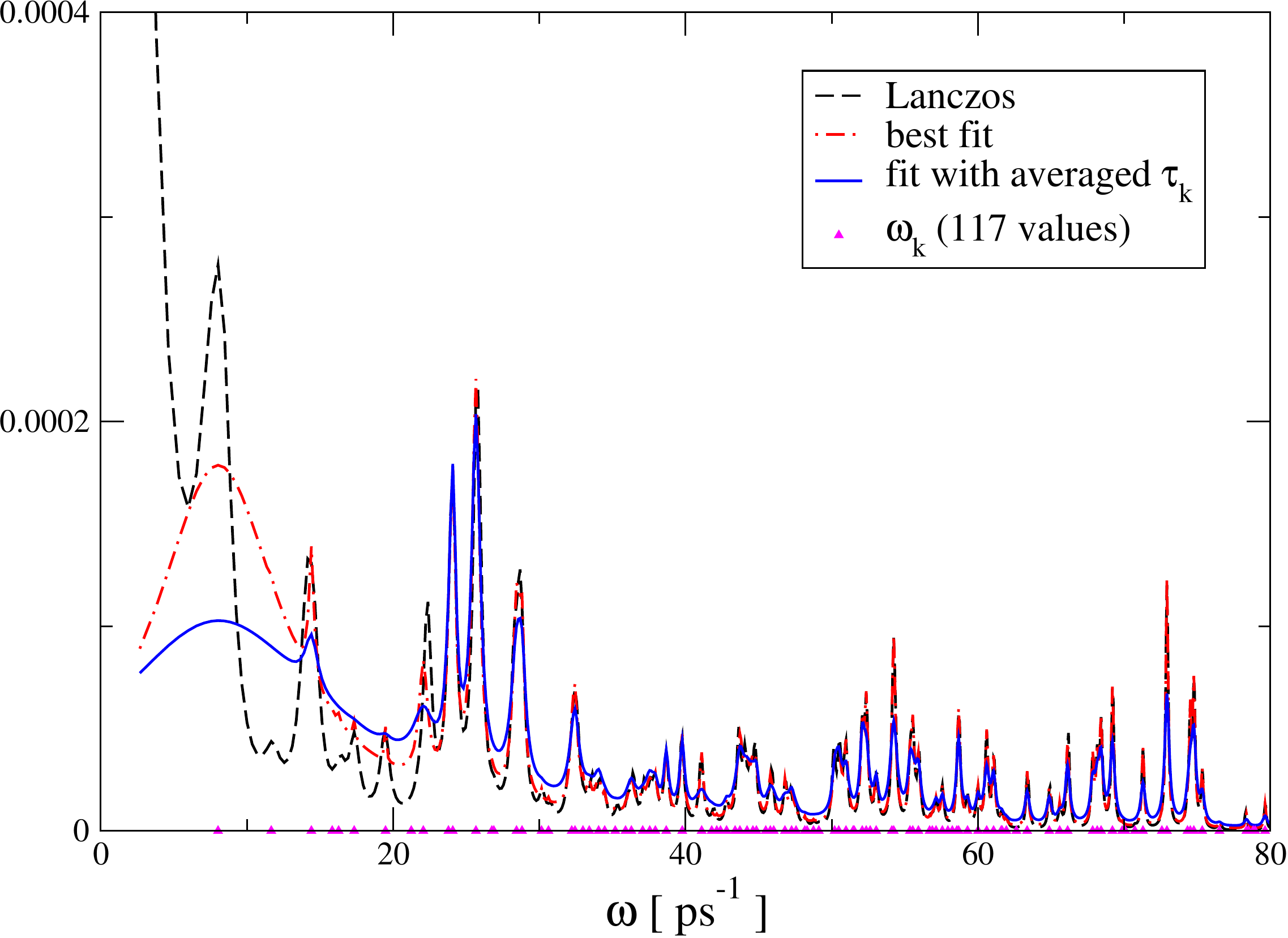}
\end{centering}
\caption{(Colour online) Typical example for the fit of a diagonal element of $\Pi_{b,b}(\omega)$
performed by using 33 different values for the vDOF peak positions $\omega_k$ (top panel)
and 117 values for $\omega_k$ (bottom panel). {As one would expect for any fitting procedure, the more}
elementary functions are put in the fit, here Lorentzian of width $1/\tau_k$ and position $\omega_k$,
the better the fit is.}
\label{fig:PIbb_fit_example}
\end{figure}

We chose to consider below the $\Pi_{b,b}(\omega)$ functions which present a lot of peaks,
as opposed to low features functions obtained with a large bath region (see Fig.~\ref{fig:PIbb_vs_bathsize}).
We do this in order to test the robustness of our fitting procedure.

Figure~\ref{fig:PIbb_fit_example} shows a typical example of our mapping procedure
for a diagonal element of $\Pi_{b,b}(\omega)$.
The best fit is given by the red curves. After fitting all the diagonal elements
$\Pi_{b,b}(\omega)$, we calculated (as explained in Sec.~\ref{sec:mapping}) an effective $\tau_k$ 
value associated with each peak at $\omega_k$, as the extended Langevin dynamics deals
with  $\{\tau_k,\omega_k\}$ parameters independent of the bath index $b$.
Using the smallest of all $\tau_k$ (for each peak at $\omega_k$), we still obtain a good
fit (blue curves) of the original $\Pi_{b,b}(\omega)$ result.

Note that as expected for any fitting procedure, the more elementary functions (Lorentzian of width 
$1/\tau_k$ and position $\omega_k$) are used for the mapping, the better the fit is. However, we
show below that both sets of fitting parameters will lead to a proper physical behaviour
of the system, i.e. as far as the thermalisation of the kinetic energy and velocity distributions
are concerned.

\subsection{Fitting the off-diagonal elements of $\Pi_{b,b'}(\omega)$}
\label{sec:fitPIbbp}

\begin{figure}
\begin{centering}
\includegraphics[width=80mm]{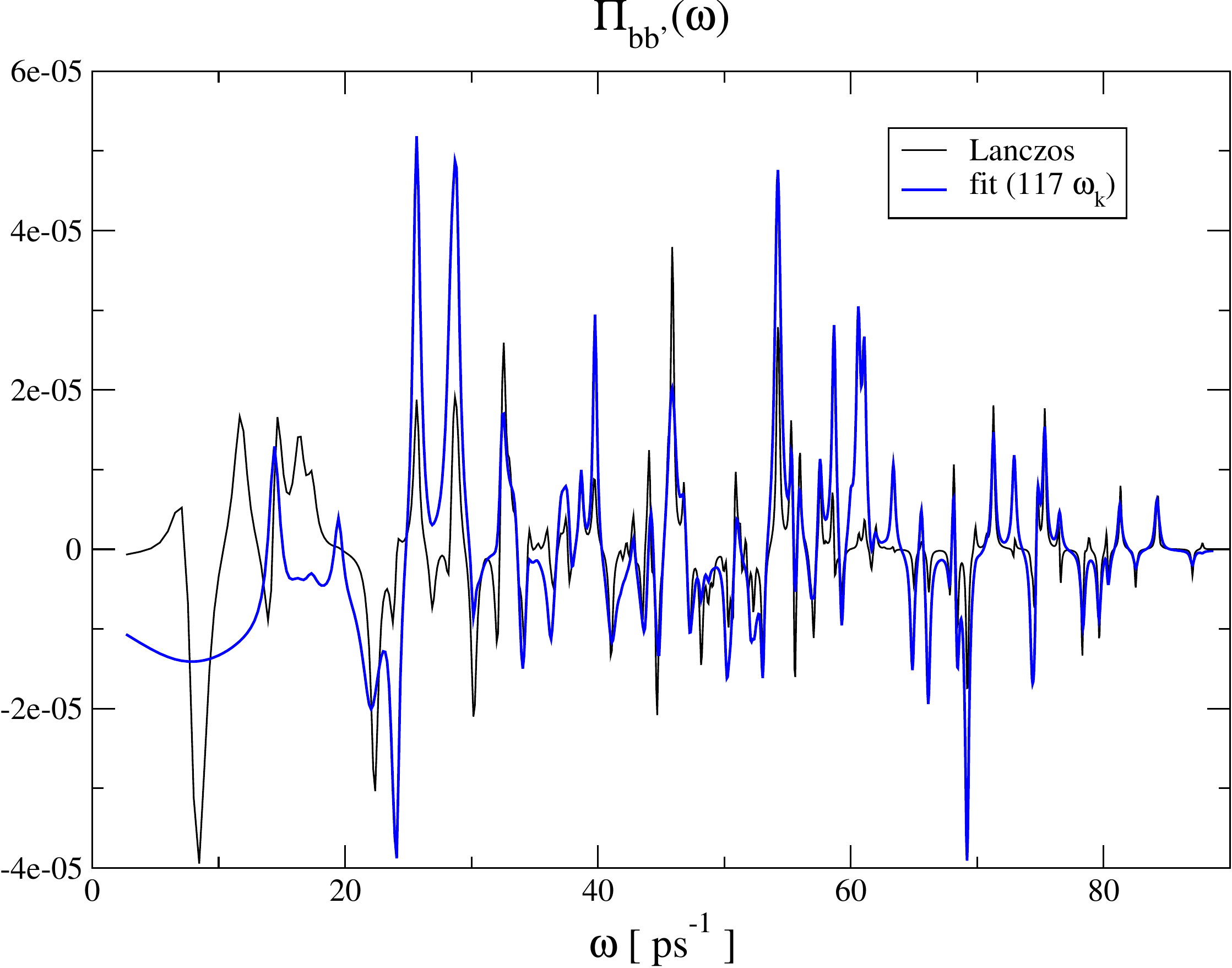}
\end{centering}
\caption{(Colour online) Typical example for the fit of an off-diagonal element of $\Pi_{b,b'}(\omega)$.
Fit is performed by using 117 different values for the vDOF peak position $\omega_k$.}
\label{fig:PIbbp_fit_example}
\end{figure}

As explained in Sec.~\ref{sec:mapping}, once the parameters  $\tau_k$ and $\vert c_b^{(k)}\vert$ are
obtained from the fits of the diagonal elements $\Pi_{b,b}(\omega)$, the proper sign of all the 
coefficients $c_b^{(k)}$ is determined from the best fit of the off-diagonal elements $\Pi_{b,b'}(\omega)$.
A typical best fit result is shown in Figure~\ref{fig:PIbbp_fit_example}.

With such a procedure, we obtain an approximate fit of the $\Pi_{b,b'}(\omega)$ function, which is 
not as good as for the diagonal elements.
However, in some ranges of frequency, the off-diagonal matrix elements are very well reproduced 
by our mapping scheme as shown in Figure~\ref{fig:PIbbp_fit_example}.

We would like to stress again that the fitting scheme of all $\Pi_{b,b'}(\omega)$ functions 
is a highly non-trivial multi-variable optimisation problem, which includes strong constraints 
(i.e. the parameters $\{\tau_k,\omega_k\}$ are independent of the bath indexes $b,b'$). 
In this paper, we have provided one possible scheme to perform such a mapping, 
but many more are available. We are currently investigating other routes \cite{SummerStudents:2014}.

\section{Results for the GLE in the extended phase space}
\label{sec:gle_ex}

\subsection{Thermalisation of the {system}}
\label{sec:Ekin}

First of all, we study how the system thermalises in our model of a realistic bath
characterised by a set of parameters $\{\tau_k, \omega_k, c_b^{(k)} \}$.
Initially, the atomic positions in the central system are at equilibrium and all velocities are
set to zero. {We then run different extended GLE dynamics simulations
using the algorithm described in detail in Appendix \ref{app:algo}.}

We want to stress that all the dynamics we have obtained, for the different sets
of parameters $\{\tau_k, \omega_k, c_b^{(k)} \}$, are stable.
We do not obtain any pathological behaviour in the calculations
of the atomic positions and velocities over thousands of time steps (runs of up to 80 ps
using a time step of $\Delta t = 1$ fs).
{In the following, we present a few selected results from all the calculations we
have performed.}

\begin{figure}
\begin{centering}
\includegraphics[width=70mm]{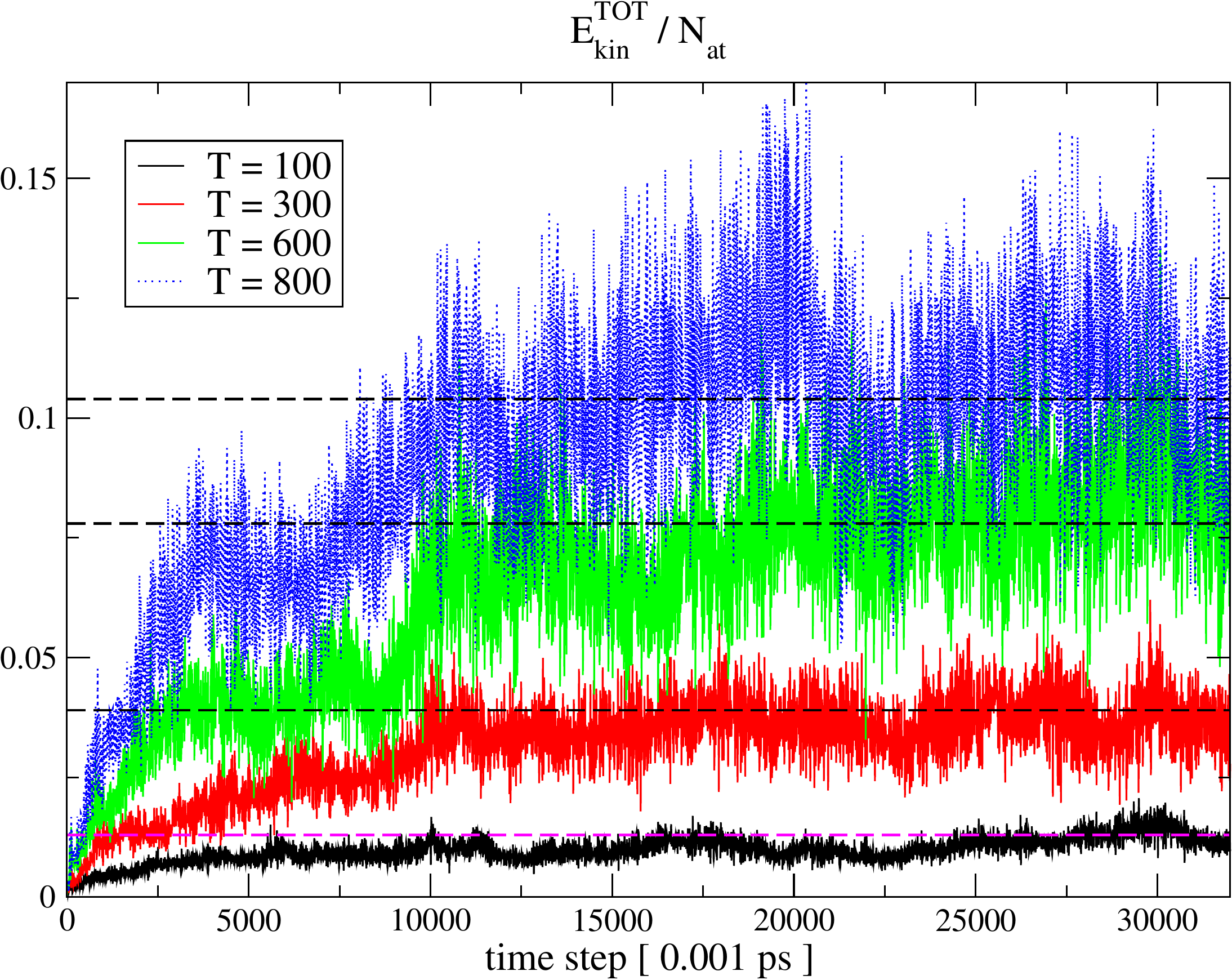}
\end{centering}
\caption{(Colour online) Total kinetic energy of the system region containing 19 atoms 
shown in {the right panel of} Fig.~\ref{fig:system1}. 
The GLE calculations are performed for different bath temperatures $T$ (in K) and for a set of fitting
parameters obtained with 33 vDOF, and for the bath region of radius
$R=7.5$ \AA\ (135 atoms), see Fig.~\ref{fig:PIbb_fit_example}.
The system thermalises to the proper equilibrium temperature after $t \sim 15-18$ ps. 
The horizontal lines correspond to the different values of $3/2 k_B T$ and show that the 
GLE dynamics properly equilibrates the system region according to the equipartition principle.
The energies are given in [eV].}
\label{fig:EkinTOT_33vDOF}
\end{figure}

\begin{figure}
\begin{centering}
\includegraphics[width=70mm]{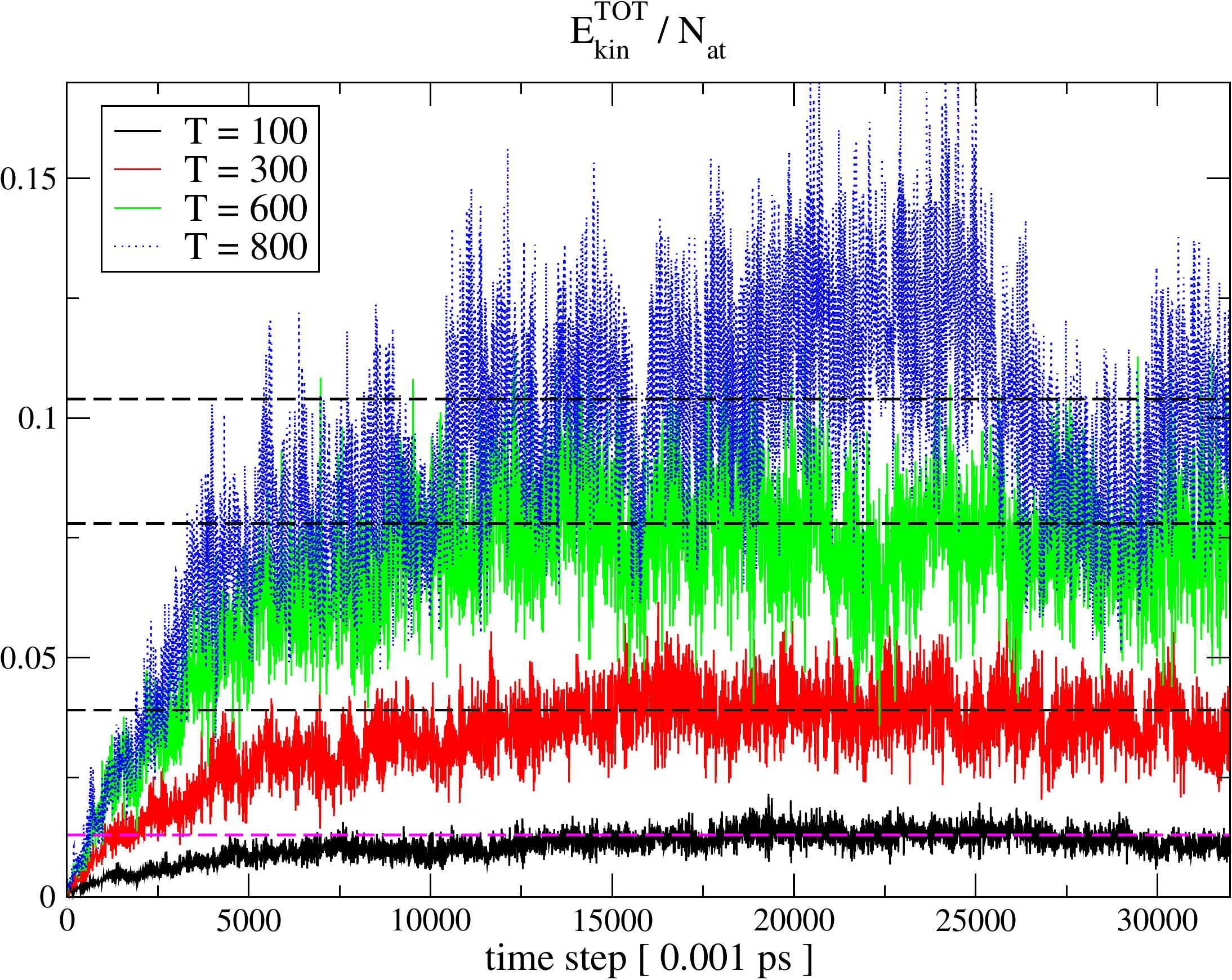}
\end{centering}
\caption{(Colour online) Total kinetic energy of the system region containing 19 atoms 
shown in {the right panel of} Fig.~\ref{fig:system1}. 
The GLE calculations are performed for different bath temperatures $T$ (in K) and for the set of fitting
parameters obtained with 117 vDOF (see Fig.~\ref{fig:PIbb_fit_example} and Fig.~\ref{fig:PIbbp_fit_example}).
The results are quantitatively different from the calculations performed with 33 vDOF
(Fig.~\ref{fig:EkinTOT_33vDOF}), but are qualitatively similar.
The system thermalises, as expected, to the proper equilibrium temperature after a shorter 
time $t \sim 12-15$ ps in comparison with Fig.~\ref{fig:EkinTOT_33vDOF}.}
\label{fig:EkinTOT_117vDOF}
\end{figure}

Figures \ref{fig:EkinTOT_33vDOF} and \ref{fig:EkinTOT_117vDOF} {represent} the evolution of
the total kinetic energy
for the system shown in {the right panel of} Fig.~\ref{fig:system1}. 
The system on which the GLE is performed contains $N_{\rm at}=19$ atoms, 
and the bath reduced region contains 68 atoms. 
The mapping of the $\Pi_{bb'}(\omega)$ functions is performed by
using 33 vDOF (see Fig.~\ref{fig:EkinTOT_33vDOF}) and 117 vDOF (see Fig.~\ref{fig:EkinTOT_117vDOF}).
We recall that during the mapping procedure, the dynamical matrix is obtained for
a bath region of radius $R=7.5$ \AA\ which contains 135 atoms {(see left panel of Fig.~\ref{fig:system1})}.

The results of our GLE calculations show that the system thermalises towards the proper equilibrium
temperature as expected, since the averaged total kinetic energy follows the equipartition principle
and oscillates around the expected value of $E_{\rm kin}^{\rm TOT} = 3/2 N_{\rm at} k_B T$.
Such a behaviour is obtained for all the temperatures $T=100,300,600,800$ K we have considered and
for different sets of fitting parameters.
The time taken by the system to reach the thermal equilibrium depends strongly on the values of
the fitting parameters, more specifically on the relaxation times $\{\tau_k \}$ associated with
the vDOF.

Further examples for the thermalisation of the system are provided in Appendix \ref{app:therma}.

\subsection{Velocity distributions}
\label{sec:velocdistrib}

\begin{figure}
\begin{centering}
\includegraphics[width=70mm]{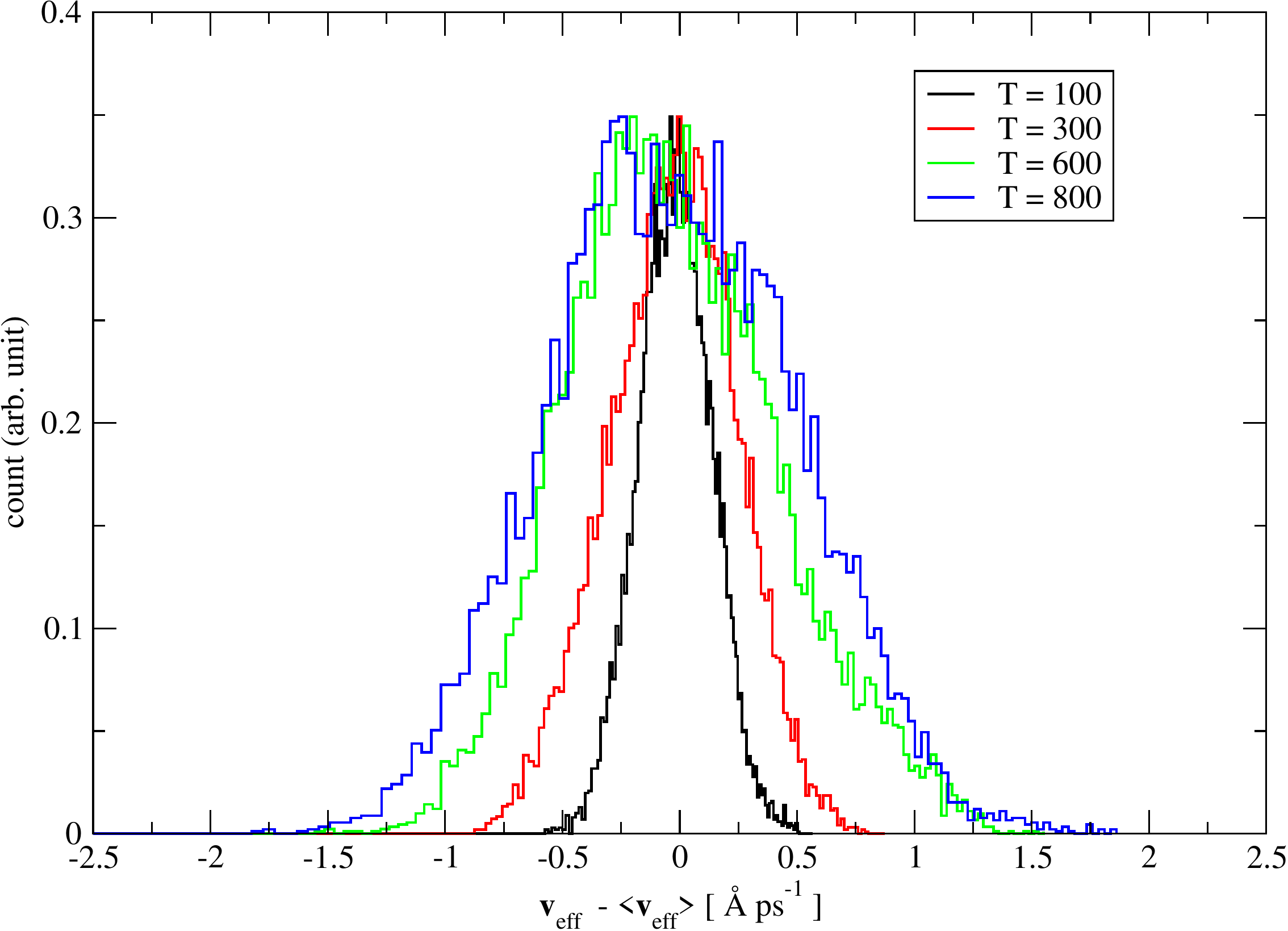}
\end{centering}
\caption{(Colour online) Histograms of the effective velocity built from the time series of the kinetic energy
shown in Fig.~\ref{fig:EkinTOT_117vDOF}. The histograms are centred around the corresponding
mean effective velocity $\langle v_{\rm eff}\rangle$. The widths of the distributions 
follow a Gaussian distribution in $\exp(-\beta m v_{\rm eff}^2/2)$.}
\label{fig:histo_veloc_from_Ekin_vs_T}
\end{figure}

From the time evolution of the total kinetic energy, we can extract an effective velocity
$v_{\rm eff}$ from the relation $1/2  m v_{\rm eff}^2 = E_{\rm kin}^{\rm TOT} / N_{\rm at}$.
Using the time series of such a velocity, we can build up a histogram of the
velocity in a range of the time span $[t_1,t_2]$ for which the system is thermalised.
Figure~\ref{fig:histo_veloc_from_Ekin_vs_T} represents such a histogram for different
temperatures, using the values of the total kinetic energy shown in
Fig.~\ref{fig:EkinTOT_117vDOF} and for the range $[t_1,t_2] = [20,32]$ ps.

We have checked that the full width at half maximum (FWHM) follows the
behaviour of a Gaussian distribution in $e^{-\beta m v_{\rm eff}^2 / 2}$, i.e. the
ratio between two FWHMs for two different temperatures is like
$\Delta v_{\rm eff}(T_1) / \Delta v_{\rm eff}(T_2) = \sqrt{T_1 / T_2}$.
In other words, such a result can be understood as follows: the system thermalises 
to the expected bath temperature, and the 
corresponding effective temperature fluctuates around the mean value 
according to a Gaussian distribution.
 
More importantly, we can also study the statistics of the velocity of individual atoms in
the central region.
For that, we build the velocity distribution $P_v(t_i)$ of the velocities $v_i=(\sum_{\alpha=x,y,z} v_{i\alpha}^2)^{1/2}$ 
of each individual atom $i$ in the central region for the set of velocities obtained at time $t_i$ 
when the system is thermalised.
In order to obtain a better statistical representation of such a distribution, we calculate
an averaged distribution 
\begin{equation}
\bar{P}_v = \sum_{i=1}^{N_{\rm ts}} P_v(t_i) / N_{\rm ts}
\end{equation}
over a set of $N_{\rm ts}$ different times $t_i$ in the time range 
$[t_1,t_2]$ for which the system is thermalised.
 
\begin{figure}
\begin{centering}
\includegraphics[width=70mm]{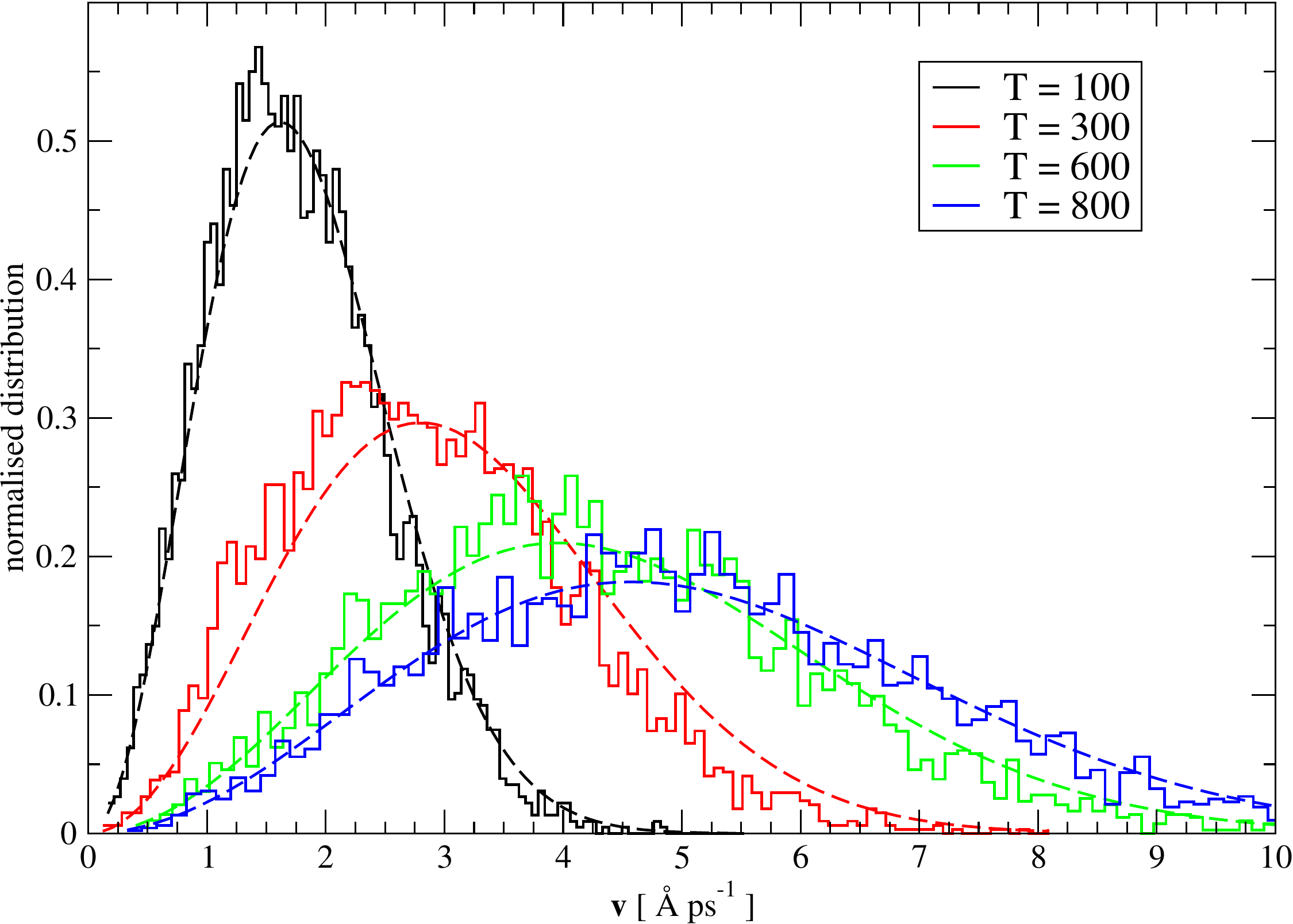}
\end{centering}
\caption{(Colour online) Histograms of the velocity distribution $\bar{P}_v$ calculated from a GLE dynamics
based on the use of 117 vDOF. $\bar{P}_v$ is obtained from $N_{\rm ts}$=220 different 
timesteps $t_i$ taken in the range $t=[30,52]$ ps.
The broken curves correspond to the Maxwell-Botzmann distribution $f_v$ and represent an
almost perfect fit between the two velocity distributions.}
\label{histo_distrib_veloc_117vDOF_vsT}
\end{figure}

An example of the velocity distribution $\bar{P}_v$ is shown in Figure~\ref{histo_distrib_veloc_117vDOF_vsT}.
The GLE calculations were performed by using the set of parameters $\{\tau_k, \omega_k, c_b^{(k)} \}$  
based on 117 vDOF. In the calculation of $\bar{P}_v$, we used $N_{\rm ts}$=220 different time steps $t_i$ 
equally spaced in the time range $t=[30,52]$ ps. 
We also compared the calculated distribution $\bar{P}_v$
with the corresponding Maxwell-Boltzmann distribution defined as
\begin{equation}
f_v = \sqrt{ \left( \frac{m\beta}{2\pi} \right)^3} 4\pi v^2 e^{-mv^2/2\beta}. 
\end{equation}

From Figure~\ref{histo_distrib_veloc_117vDOF_vsT}, we can see an almost perfect match between
the two distributions $\bar{P}_v$ and $f_v$.

To conclude this section, we can confidently say that our extended GLE calculations provide
a good thermostat model, in the sense that the central system thermalises towards the 
expected temperature, with expected Gaussian fluctuations around the mean value of the
effective temperature. More importantly, the thermostat provides the correct canonical
distribution of the velocities in the central region once the system is thermalised.

\subsection{Velocity autocorrelation functions}
\label{sec:vacf}

One last dynamical quantity that we need to examine is the velocity autocorrelation functions
of the central system.
The velocity autocorrelation functions (VACF) are calculated from
\begin{equation}
\langle v(t_0) v(t+t_0)\rangle = \sum_{i\alpha} v_{i\alpha}(t_0) v_{i\alpha}(t+t_0) / (3 N_{\rm at}) 
\end{equation}
for all atoms $i$ of the central region and with $t>t_0$.

For the two times $t_0$ and $t$ being within the time range where the system is thermalised,
the VACF should be dependent only on the time argument difference $\Delta t = (t+t_0) - t_0$,
i.e. independent of the initial time $t_0$.
In order to obtain a better statistical representation of the VACF, we also
calculate an averaged VACF from {different $N_{\rm samp}$ samplings
of the initial time $t_0$ in a time range where the system is thermalised}:
\begin{equation}
\overline{\langle v(0) v(t)\rangle} = \sum_{\{t_0\}}^{N_{\rm samp}}  \langle v(t_0) v(t+t_0)\rangle / N_{\rm samp} .
\end{equation}

\begin{figure}
\begin{centering}
\includegraphics[width=70mm]{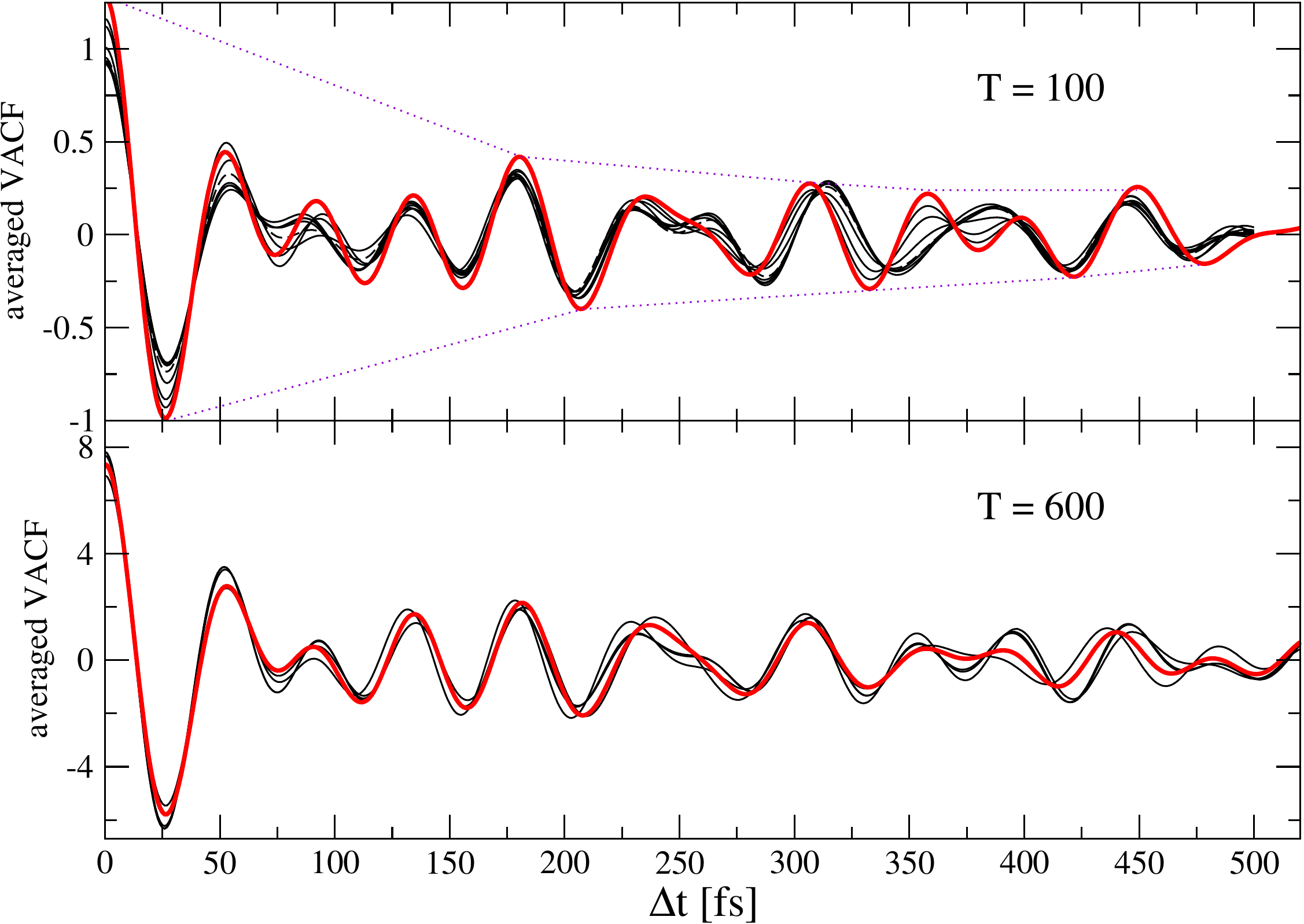}
\end{centering}
\caption{(Colour online) Velocity autocorrelation functions for the central system region containing 19 atoms. 
Calculations for the averaged VACF $\overline{\langle v(0) v(t)\rangle}$
are obtained with the set of fitting parameters corresponding to Fig.~\ref{fig:EkinTOT_117vDOF}
and for two temperatures $T=100$ K (top panel) and $T=600$ K (bottom panel).
The average is obtained from 300 different samplings of the initial time $t_0$ 
over the time range $[40,41.5]$ ps. The different thin lines correspond to
different sets of sampling. The dotted lines in the upper panel
are a guide for the eye to show the decaying
of the VACF with the time difference $\Delta t$.}
\label{fig:VACF_117vDOF_T100}
\end{figure}

Figure~\ref{fig:VACF_117vDOF_T100} represents the corresponding averaged
velocity autocorrelation functions $\overline{\langle v(0) v(t)\rangle}$ 
for the central system containing 19 atoms shown in Fig.~\ref{fig:system1} and for the 
temperature $T=100$ K.
The GLE calculations were performed with the set of fitting parameters based on 117 vDOF. 
The averaged VACF was calculated for $t_0 \sim 40$ ps and $t \sim 40.5$ ps and using 
300 different samplings of the initial time $t_0$ over the time range $\sim[40,41.5]$ ps.
Our GLE results show the proper decaying behaviour of the VACF with the time difference $\Delta t$.
It is interesting to note that the loss of the velocity correlation occurs on a much shorter time 
scale than the time scale corresponding to the thermalisation of the system
(starting from zero velocities).

\subsection{Simplified Langevin dynamics with a single friction coefficient}
\label{sec:LangevinGaussian}

To further confirm the validity of our approach, we now compare our GLE results
with the more conventional approach of the Langevin dynamics, using a more
heuristic description of the dissipation in the system:

\begin{equation}
\begin{split}
\dot{\mathbf{p}} = - \mathbf{\nabla}_\mathbf{r} \bar{V}(\mathbf{r}) - \gamma \mathbf{p} + \boldsymbol{\xi}_G
\end{split}
\label{eq:normalLangevin}
\end{equation}
with the momentum vector $\mathbf{p} = \mathbf{M}  \dot{\mathbf{r}}$ and the random noise
vector $\boldsymbol{\xi}_G$. 
The latter follows a Gaussian distribution \cite{Uhlenbeck:1930,Gillespie:1996}. The random noise has the dispersion
which is related directly to the friction coefficient via the well-known expression
$\sigma_i^2 = 2 M_i \gamma k_B T / \Delta t$, where $\Delta t$ is the time step of the dynamics.
{Note that the friction and random forces are applied here to all the atoms of the central
system.}
The Gaussian Langevin dynamics has already been implemented in the code 
LAMMPS \cite{Toton:2010,Plimpton:1995}.

\begin{figure}
\begin{centering}
\includegraphics[width=70mm]{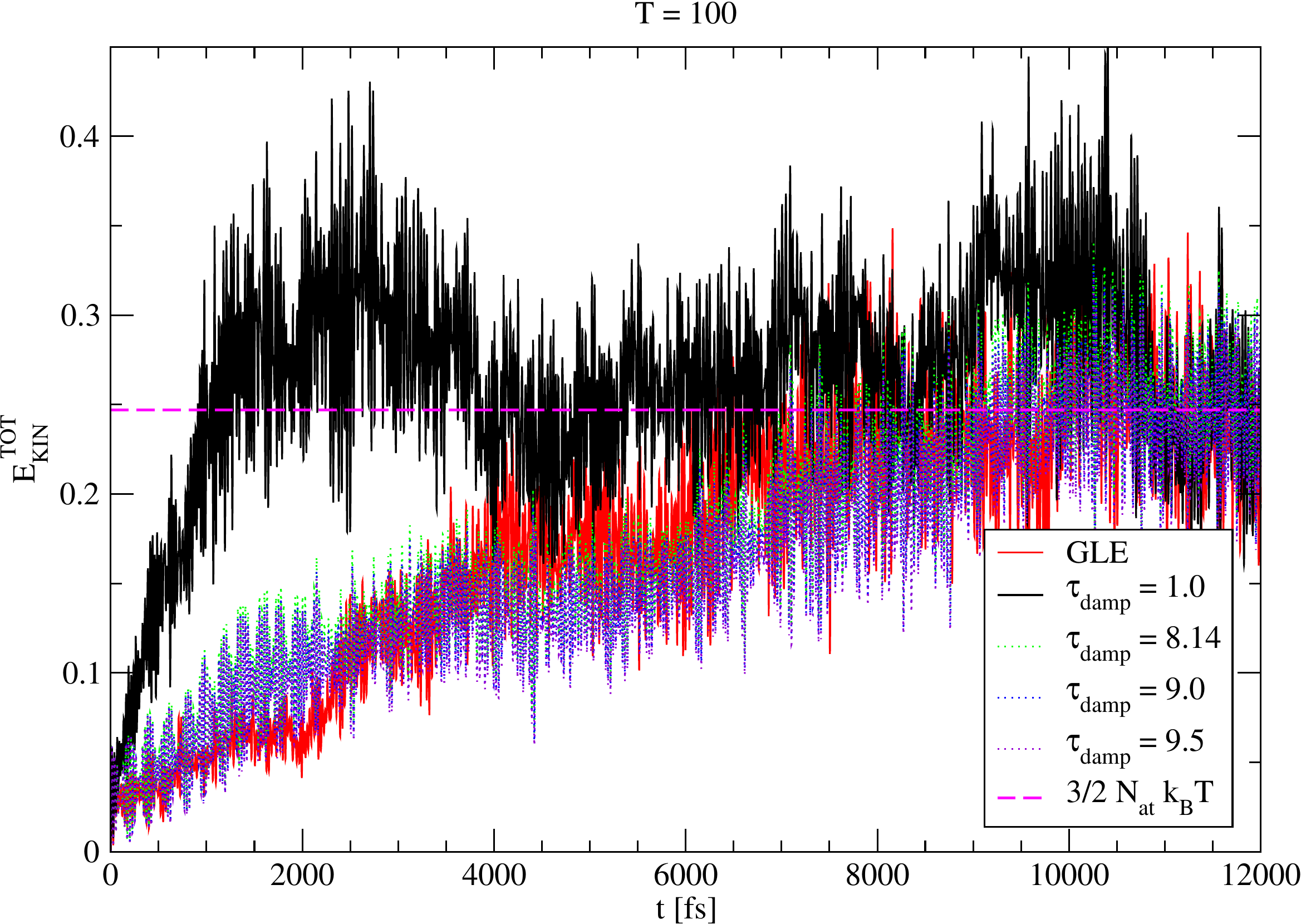}
\end{centering}
\caption{(Colour online) Total kinetic energy of the system region containing 19 atoms
shown in {the right panel of} Fig.~\ref{fig:system1}. 
{The plots show a} comparison between GLE calculations and conventional Langevin dynamics with a simple
friction constant $\gamma$ for bath temperature of $T = 100$ K.
All total kinetic energies converge towards the expected thermodynamical equilibrium value.
One obtains a good correspondence between the conventional Langevin dynamics and the GLE
dynamics for a friction constant $\gamma = 1/\tau_{\rm damp}$, with $ \tau_{\rm damp}\sim  9.0 - 9.5$  ps.
Initially, all velocities are set to zero.}
\label{fig:EkinTOT_LangGauss_GLE_T100_117vDOF_Tinit0}
\end{figure}

Figure~\ref{fig:EkinTOT_LangGauss_GLE_T100_117vDOF_Tinit0} shows the time evolution of the total kinetic 
energy of the system region containing 19 atoms ({right panel of} Fig.~\ref{fig:system1}). 
Both GLE and conventional Langevin dynamics provide a total kinetic energy that converges towards the 
expected thermodynamical equilibrium value of $3/2 N_{\rm at} k_B T$ (with $T=100$ K).
One can see that the conventional Langevin dynamics results can fit fairly well the results obtained
from the GLE calculations by adjusting the friction coefficient $\gamma$.
For the target temperature of the bath $T=100$ K and the initial temperature $T_{\rm init} =0$
(initially, all velocities are set to zero), we obtain the best correspondence between the 
conventional Langevin dynamics and the GLE dynamics for the friction constant value 
$\gamma=1/\tau_{\rm damp}$ with $\tau_{\rm damp} \sim  9.0 - 9.5$ ps.

Such a range of values for the friction constant of the conventional Langevin dynamics 
seems to provide the appropriate behaviour of the total kinetic energy for the model
bath we have used.
We have checked that the range $ \tau_{\rm damp}\sim  9.0 - 9.5$ ps provides
the appropriate behaviour of $E_{\rm kin}^{\rm TOT}$ when the dynamics are started
with initial velocities different from zero.
Furthermore, we have also checked that such a range of  $\tau_{\rm damp}$ is
appropriate for a range of temperatures going from $T=100$ to $T=600$ K.

\begin{figure}
\begin{centering}
\includegraphics[width=70mm]{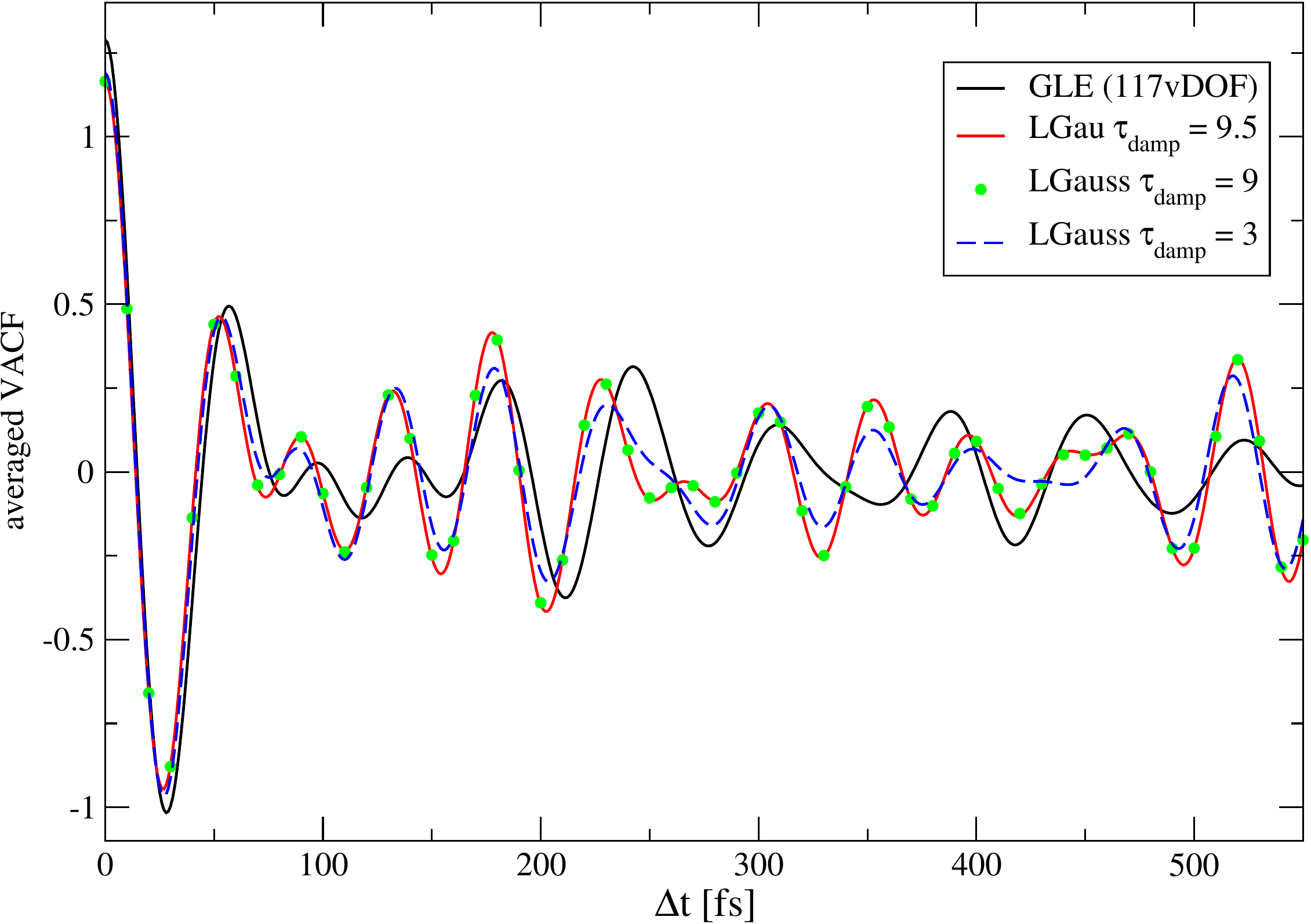}
\end{centering}
\caption{(Colour online)
Velocity autocorrelation functions for the central system region containing 19 atoms. 
{The plots show the averaged VACF $\overline{\langle v(0) v(t)\rangle}$
for a temperature $T=100$ K for the GLE runs based on 117 vDOF and for the
conventional Langevin dynamics (LGauss) with 
different friction constants $\gamma = 1/\tau_{\rm damp}$.}
The average is obtained from 600 different samplings of the initial time $t_0$ 
over the time range $[48.5,49.0]$ ps.
All averaged VACF are performed in exactly the same manner.}
\label{fig:VACF_LangGauss_vs_117vDOF_T100}
\end{figure}

Finally we can compare the VACF obtained from the conventional Langevin dynamics
with our GLE calculations. Figure~\ref{fig:VACF_LangGauss_vs_117vDOF_T100} shows the
averaged VACF for one temperature. The averages of the VACF are performed in exactly
the same way for all the calculations.
We can observe a good correspondence between the GLE and conventional Langevin
calculations. The loss of correlation in the velocities appears slightly earlier
for the GLE calculations. \HN{The dependence of the VACF upon the friction constant seems 
weaker
than for the kinetic energy, however the best correspondences are obtained for
the range of damping $ \tau_{\rm damp}\sim  9.0 - 9.5$ ps.}

\HNrev{It should be noted that, for the present model of a homogeneous LJ solid
used in our calculations, the results obtained with the conventional Langevin dynamics
are indeed very similar to the results obtained with our more general and complex
GLE method. However, there is one fundamental difference between the two approaches:
the conventional Langevin dynamics requires an {\it a priori} unknown input parameter,
i.e. the friction constant $\gamma$, which is not the case for our GLE approach.
As shown above, our GLE approach can be used to extract such an input parameter for
the heuristic Langevin equation.
}

\section{Summary and Discussion}
\label{sec:ccl}

In this paper, we have implemented the GLE scheme developed in 
Refs.~[\onlinecite{Kantorovich:2008}] and [\onlinecite{Stella:2014}]
and have shown several applications for systems described at the atomic level.
We recall that this GLE scheme goes beyond a {bi-linear} coupling
between the central system and the bath, and permits us to have a realistic description
(i.e. at the atomic level) of both the dissipative central system and its surrounding 
bath.
This implementation of the GLE scheme is done in the classical MD code LAMMPS.

We have shown how to obtain the vibrational properties of a realistic bath and
how to convey such properties into an extended Langevin dynamics by the use of
the mapping of the bath vibrational properties onto a set of auxiliary DOF,
see Eq.~(\ref{eq:mapping_PI_matrix}).

Different applications of such a mapping scheme and of the corresponding
extended Langevin dynamics were given for different models of a LJ solid.
In this manuscript, the implementation of our GLE method
is done for pair-wise interatomic potential. The use of such potentials makes the 
calculations of the different quantities, such as 
$f_b(\{r_{i\alpha}\})$ and $g_{i\alpha,b}(\{r_{i\alpha}\})$
to be evaluated twice at each time step, much faster. Implementation for any type of N-body
potential is under consideration.

All our calculations show that our GLE scheme provide a stable Langevin dynamics,
with the dissipative/relaxation processes properly described.
The total kinetic energy of the central system always thermalises toward the expected 
bath temperature, with appropriate fluctuation around the mean value.
More importantly, we obtain a velocity distribution for the individual atoms in
the central system which follows the expected canonical distribution at the
corresponding temperature. This confirms that both our GLE scheme and our mapping
procedure onto an extended Langevin dynamics provide the correct thermostat.
We have also examined the corresponding VACF and found that the velocities lose
correlations as expected, however the corresponding time scale is much shorter
than the time taken by the system to reach thermalization.

We have also compared our GLE results with respect to more conventional
Langevin dynamics based on a single relaxation time (i.e. single friction
coefficient). Our calculations have shown the possibility of extracting 
an effective friction coefficient from our realistic bath model, which
then could be used {\it a posteriori} in a much less expensive Langevin dynamics.
Our calcutations have shown that the obtained effective friction coefficient
is independent on the initial distribution of the velocities and on the
temperature of the system (at least for the range 100---600 K we have
considered).

\HNrev{One has to have in mind, however, that it is only for the rather 
simple model system considered here that the friction coefficient of the 
heuristic Langevin dynamics was found to be temperature independent. 
There is no reason to believe that this is a general rule and that for 
other systems, e.g. highly inhomogeneous, it will still be the case. 
Furthermore, in the cases of heterogeneous systems different values 
of the friction coefficient for different species need to be found. 
It is not clear {\it a priori} what value is to be used, and also how the 
right value can be chosen in practice. Indeed, as was shown in
 Ref.[\onlinecite{Kantorovich:2008b}], 
any value of the friction coefficient, even if applied not to all atoms 
of the system, would always bring the system to the equilibrium state 
described by the corresponding canonical distribution. Hence the value 
of the friction parameter(s) can only be obtained by running genuinely 
non-equilibrium simulations, e.g. on heat transport, rate of equilibration 
and so on. It seems that using GLE eliminates all these problems by 
providing a clear and fundamentally sound platform for either running 
(more expensive) GLE type calculations or using GLE for fitting the 
value(s) of the friction coefficient(s). If necessary, temperature dependent 
friction is also within reach.
}

Finally, we would like to comment on two different points.
First, the results presented in this paper were obtained for a homogeneous 
``rather simple'' system (i.e. made of only one chemical species),
furthermore the system does not have a complicated geometry. 
Our GLE scheme is however applicable to much more complex systems 
(i.e. highly heterogeneous, and with complex structures
like bio-like molecules deposited on rough surfaces).
The results presented in this paper should be mostly understood as a proof
of principle of our methodology. 

For complex systems, we expect that the bath vibrational properties 
will present more specific features
which will lead to more specific properties of the memory kernel.
In turn, the properties of such a kernel will strongly affect, by some
kind of selective processes, the efficiency of some vibrational modes of 
the central region to exchange energy with the surrounding bath. 
We expect that such specific bath properties will be central
in the thermalization and relaxation processes of (small to large) 
molecules grafted onto surfaces or clusters (and into the presence or 
not of solvents).

\HN{Second, a large number of equilibrium thermostats has been designed up to
date (see Refs.~[\onlinecite{Toton:2010}] and references therein).} The GLE can be used to provide exactly
the same results as obtained from these equilibrium thermostates, albeit with
a higher computational cost.
However, the main advantage of the GLE, as compared with the other available
equilibrium thermostats, is that it is also applicable to the study of 
nonequilibrium processes.
For instance, the GLE technique is, by essence, naturally applicable for studying
the phonon contribution to thermal transport through bulk materials or nano-junctions.
Such nonequilibrium processes can be treated by coupling the central system
to more than one bath.
Each bath would be at its own equilibrium, and one cannot define a single temperature 
for the whole system. 
In that case, the central system does not evolve towards an equilibrium state, but will 
eventually reach a steady state regime 
characterised by heat flows between the central system and the baths.
To study such processes, the GLE equation~(\ref{eq:GLEmatrix_form}) can be 
generalised to include the nonequilibrium conditions
when the different baths are independent (i.e. not coupled to each other in any way).
For that we simply need to extend the number of virtual DOF to obtain a set of parameters
$\{\tau_k, \omega_k, c_b^{(k)} \}_\nu$ for each bath $\nu$ at temperature $T_\nu$. 
Each bath $\nu$ will also be characterised by its own dynamical matrix and matrix elements
${\Pi}_{b,b'}^\nu$. 
The implementation of such nonequilibrium extended Langevin dynamics is
currently under development.

\begin{acknowledgements}

HN warmly thanks L. Pizzagalli for fruitful discussions and for providing important 
informations about the calculations of the dynamical matrix within LAMMPS.
We acknowledge financial support from the UK EPSRC, under Grant No. EP/J019259/1.

\end{acknowledgements}

\appendix

\section{Bath vibration propagator and $\boldsymbol{\Pi}$ matrix}
\label{app:Phproga}

In Ref.~[\onlinecite{Kantorovich:2008}] it is shown that, in the time representation,
the matrix $\boldsymbol{\Pi}(t-t')$ is related to the bath propagator $\boldsymbol{\mathcal{D}}(t-t')$
via $\partial_\tau\boldsymbol{\Pi}(\tau)=-\boldsymbol{\mathcal{D}}(\tau)$.

The bath propagator $\boldsymbol{\mathcal{D}}(t-t')$ is the solution of the harmonic dynamics of the
bath DOF
\begin{equation}
\sum_{b_1} \left[ \partial_t^2 + D_{b,b_1}\right] \mathcal{D}_{b_1,b'}(t-t') = \delta(t-t') \delta_{b,b'} ,
\label{eq:bath_prop}
\end{equation}
where $[\bm{D}]_{b,b'}$ are the elements of the dynamical matrix of the bath region.

The elements of the matrix $\boldsymbol{\Pi}$ are given by \cite{Kantorovich:2008}
\begin{equation}
{\Pi}_{b,b'}(t-t') = \sum_\lambda e_\lambda^{b\dag} e_\lambda^{b'}\ \frac{\cos\omega_\lambda(t-t')}{\omega_\lambda^2}
\label{eq:PIttprim}
\end{equation}
where $\lambda$ labels the eigenstates of the dynamical matrix $\bm{D}$ with eigenvalues
$\omega_\lambda^2$ and eigenvectors $\bm{e}_\lambda$ with component $e_\lambda^b$ in the bath region. 

As all quantities depend only on a single time argument, one can pass into the energy representation
after using the Fourier transform. 

The bath propagator $\mathcal{D}(\omega)$ is then the solution of 
\begin{equation}
\sum_{b_1}[(i\omega)^2\bm{1} + \bm{D}]_{b,b_1} \mathcal{D}_{b_1,b'}(\omega)=\delta_{b,b'}
\end{equation}
and
\begin{equation}
\begin{split}
{\Pi}_{b,b'}(\omega) & = \sum_\lambda e_\lambda^{b\dag} e_\lambda^{b'} \ \frac{1}{\omega_\lambda^2}
\left( \delta(\omega - \omega_\lambda) + \delta(\omega + \omega_\lambda) \right) \frac{2\pi}{2} \\
& = \sum_\lambda e_\lambda^{b\dag} e_\lambda^{b'} \ \frac{2\pi}{\vert\omega_\lambda\vert} \delta(\omega^2 - \omega_\lambda^2) \\
& = \frac{2\pi}{\vert\omega\vert} \sum_\lambda e_\lambda^{b\dag} e_\lambda^{b'} \ \delta(\omega^2 - \omega_\lambda^2) .
\end{split}
\label{eq:PIomega}
\end{equation}

It is now easy to find the relationship between $\boldsymbol{\Pi}(\omega)$ and $\boldsymbol{\mathcal{D}}(\omega)$:
\begin{equation}
{\Pi}_{b,b'}(\omega) = - \frac{2}{\vert\omega\vert} {\rm Im} \left[ \omega^2 \bm{1} - \bm{D} + i\varepsilon \right]^{-1}_{b,b'} ,
\label{eq:PIomegaDomega}
\end{equation}
by introducing a small imaginary part in 
\begin{equation}
\mathcal{D}_{b,b'}(\omega) = - \sum_\lambda e_\lambda^{b\dag} e_\lambda^{b'} \ (\omega^2-\omega_\lambda^2 + i\varepsilon)^{-1}
\label{eq:Dwitheta}
\end{equation}
and using the fact that
$i\omega\boldsymbol{\Pi}(\omega)=-\boldsymbol{\mathcal{D}}(\omega)$.

\section{Verlet-type algorithm for the extended Langevin dynamics}
\label{app:algo}

Following the prescriptions given in Ref.~[\onlinecite{Stella:2014}], we use the following algorithm for a
single time-step $\Delta t$.
The algorithm is derived, in a Verlet-style, from a different splitting and a Trotter-like decomposition of the 
total Liouvillian for the extended Langevin dynamics of the system DOF, $r_{i\alpha}$, and the virtual DOF
$s_{1,2}^{(k)}$.
{Such a decomposition has been shown to provide a more appropriate description of the velocity correlation 
functions \cite{Leimkuhler:2013}.}

{Algorithm:}
\begin{equation}
\begin{split}
& \text{(A) Randomise and propagate the vDOF} \\
& s_{x}^{(k)} \leftarrow a_k s_x^{(k)} + b_k \xi_x^{(k)} \\
& \text{(B) Calculate all $f_b(\{r_{i\alpha}\})$ and $g_{i\alpha,b}(\{r_{i\alpha}\})$} \\
& \text{(C) Propagate the DOF and vDOF} \\
& v_{i\alpha} \leftarrow v_{i\alpha} + \left( f_{i\alpha} + f_{i\alpha}^{\rm pol} + f_{i\alpha}^{p{\rm GLE}}\right) \frac{\Delta t}{2 m_i}\\
& s_{2}^{(k)} \leftarrow s_{2}^{(k)}-\omega_{k}s_{1}^{(k)}\frac{\Delta t}{2}\\
& r_{i\alpha} \leftarrow r_{i\alpha}+v_{i\alpha} \Delta t\\
& \text{(D) Recalculate all $f_b(\{r_{i\alpha}\})$ and $g_{i\alpha,b}(\{r_{i\alpha}\})$} \\
& \text{(E) Propagate the DOF and vDOF} \\
& s_{1}^{(k)} \leftarrow s_{1}^{(k)}+\left(\omega_{k}s_{2}^{(k)} + f_k^{s{\rm GLE}}\right)\Delta t\\
& v_{i\alpha} \leftarrow v_{i\alpha} + \left( f_{i\alpha} + f_{i\alpha}^{\rm pol} + f_{i\alpha}^{p{\rm GLE}}\right) \frac{\Delta t}{2 m_i}\\
& s_{2}^{(k)} \leftarrow s_{2}^{(k)}-\omega_{k}s_{1}^{(k)}\frac{\Delta t}{2}\\
& \text{(F) Randomise and propagate the vDOF} \\
& s_{x}^{(k)} \leftarrow a_{k}s_{x}^{(k)}+b_{k}\xi_{x}^{(k)}
\end{split}
\label{eq:algo}
\end{equation}
where the different forces, $f_{i\alpha}, f_{i\alpha}^{\rm pol},f_{i\alpha}^{p{\rm GLE}}, f_k^{s{\rm GLE}}$ are explained below.
The force 
\begin{equation}
f_{i\alpha} = -\frac{\partial {V}(\mathbf{r})}{\partial r_{i\alpha}}
\label{eq:f}
\end{equation}
is the force acting on the system DOF ${i\alpha}$
due to the interaction between the atoms in the system and in the bath region(s);
the ``polaronic'' force $f_{i\alpha}^{\rm pol}$
\begin{equation}
\begin{split}
f_{i\alpha}^{\rm pol} & = \sum_{b,b'} \sqrt{\mu_l \mu_{l'}}\ g_{i\alpha,b}\ \Pi_{bb'}(0) f_{b'} \\
	              & = \sum_{b,b',k} \sqrt{\mu_l \mu_{l'}}\ g_{i\alpha,b}(\{r_{i\alpha}\})\ c_b^{(k)} c_{b'}^{(k)}\ f_{b'}(\{r_{i\alpha}\})
\end{split}
\label{eq:fpol}
\end{equation}
(with $b \equiv l\gamma$ for the bath DOF)
is the force acting on the system DOF ${i\alpha}$ due to the interaction between the system and bath regions
which induces a displacement of the positions of the harmonic oscillators characterising the bath.
In Eq.~(\ref{eq:fpol}), we used the fact that $\Pi_{bb'}(0)$ is the inverse Fourier transform (evaluated at $\tau=0$) 
of $\Pi_{bb'}(\omega)$ given by Eq.~(\ref{eq:mapping_PI_matrix}).

The force $f_{i\alpha}^{p{\rm GLE}}$ acts on the system DOF ${i\alpha}$ and arises from the generalised Langevin
equations:
\begin{equation}
f_{i\alpha}^{p{\rm GLE}} = \sum_{b,k}\sqrt{\frac{\mu_l}{\bar\mu}}\ g_{i\alpha,b}(\{r_{i\alpha}\})\ c_{b}^{(k)} s_{1}^{(k)}
\label{eq:fGLEp}
\end{equation}
and the force $f^{s{\rm GLE}}$ acts on the vDOF $s_1^{(k)}$ and also arises from the generalised Langevin
equations
\begin{equation}
f_k^{s{\rm GLE}} = - \sum_{i\alpha,b} \sqrt{\mu_l \bar\mu}\ g_{i\alpha,b}(\{r_{i\alpha}\})\ c_{b}^{(k)} v_{i\alpha}
\label{eq:fGLEs}
\end{equation}

The integration of the dissipative part of the dynamics of the vDOF (see steps (A) and (F) in the algorithm) 
includes the coefficients $a_k={\rm exp}(-\Delta t / 2 \tau_k)$ and $b_k=[k_BT \bar\mu ( 1 - a_k^2) ]^{1/2}$
and the uncorrelated random variable $\xi_{1,2}^{(k)}$ corresponding to the white noise.

\section{Further examples for the system thermalisation}
\label{app:therma}

\begin{figure}
\begin{centering}
\includegraphics[width=70mm]{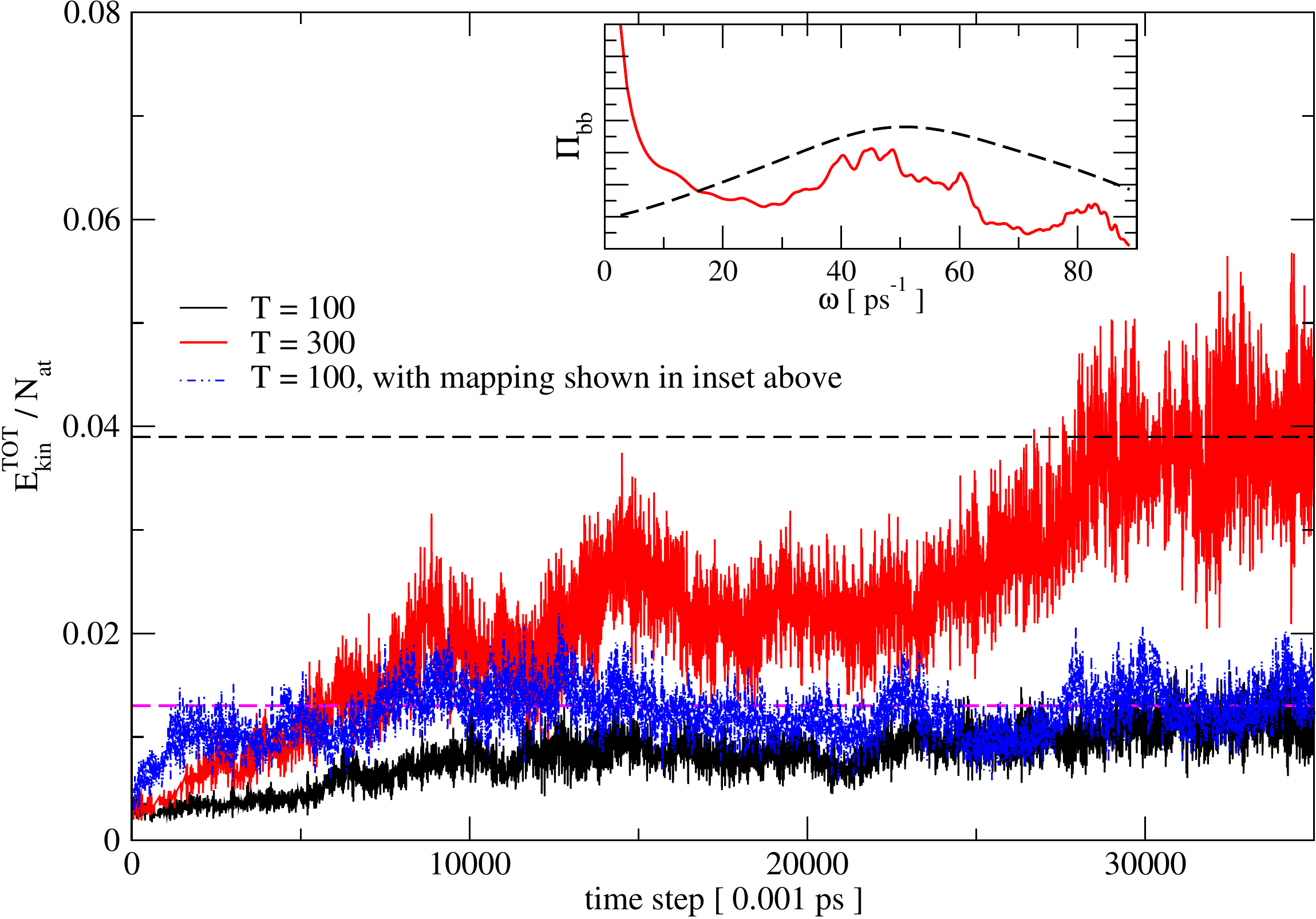}
\end{centering}
\caption{(Colour online) Total kinetic energy of the system region containing 19 atoms 
shown in {the right panel of} Fig.~\ref{fig:system1}. 
The GLE calculations are performed for different bath temperatures $T$ (in K) and for the set of fitting
parameters obtained with 48 vDOF, and for a larger bath region of radius $R=12.5$ \AA\ (555 atoms) 
(see a corresponding $\Pi_{b,b}$ function given by the red curve in Fig.~\ref{fig:PIbb_vs_bathsize}).
The system thermalises, as expected, to the proper equilibrium temperature after a time $t \sim 30$ ps.
\HN{The inset represents a poor fitting of the $\Pi_{bb'}$ functions with 27 vDOF, with the dashed curve
representing the fit and the red curve the exact result.}
The corresponding GLE calculations for the total kinetic energy (blue broken line) are shown for $T=100$.
The system thermalises much faster because the corresponding values of the $\tau_k$ parameters
are much smaller than for the fit using 48 vDOF.}
\label{fig:EkinTOT_sphere3.2_different_fit}
\end{figure}

Figure \ref{fig:EkinTOT_sphere3.2_different_fit} shows another example of the time evolution of the total
kinetic energy of the central system.
For these calculations, the set of parameters $\{\tau_k, \omega_k, c_b^{(k)} \}$ corresponds to 48 vDOF, and
the mapping is performed for the larger bath region of radius $R=12.5$ \AA\ (555 atoms) (for an example
of the corresponding $\Pi_{bb}$ functions see the red curve in Fig.~\ref{fig:PIbb_vs_bathsize}).
Once more, we can observe the thermalization of the system towards the expected equilibrium thermodynamical
values for the two different temperatures. However the overall dynamics is slower than in the previous
two cases. Such a behaviour depends on the values of the parameters $\{\tau_k, \omega_k, c_b^{(k)} \}$
obtained from the mapping.

We would like to mention that we can perform an analysis of the temporal evolution of kinetic
energy in terms of the values of the relaxation times.
Such an analysis is approximate, but still good enough when the {spreading} of the different
values of the parameters $\tau_k$, for a given fit, is not too large. 
In such a case, all $\tau_k$ values are almost the same. For the example of a poor fit shown in 
the inset of Fig.~\ref{fig:EkinTOT_sphere3.2_different_fit}, we have $\tau_k \sim 0.05$ ps for all
the 27 vDOF. The corresponding total kinetic energy 
(blue curve in Fig.~\ref{fig:EkinTOT_sphere3.2_different_fit}) 
approaches the thermal equilibrium value more quickly than for the mapping obtained with 48 vDOF.
Indeed, for the mapping done with 48 vDOF, we have $\tau_k$ parameters with more spread values
and an averaged relaxation time is around 2-3 ps which is much larger than $\sim 0.05$ ps
and explains why the system (described with the 48 vDOF) thermalises on a longer time-scale
than the system described by a poor fit with 27 DOF.

For the results presented in Fig.~\ref{fig:EkinTOT_33vDOF} and Fig.~\ref{fig:EkinTOT_117vDOF}, 
the distribution of the values of the parameters $\tau_k$ is substantially broader with values 
ranging from $\tau_k \sim 0.06$ to $\sim 6$ ps for the mapping made with 117 vDOF, 
and from  $\sim 0.06$ to $\sim 14$ ps for the mapping made with 33 vDOF. 
Correspondingly, the time taken by the system to thermalize is intermediate between the 
thermalization times shown in Figure \ref{fig:EkinTOT_sphere3.2_different_fit}.



\begin{thebibliography}{55}
\bibitem{note:1}
{These parameters are taken from the literature to mimic copper.  We are well aware 
that the LJ potential is not fully appropriate to describe metallic systems, 
however, as we have mentioned in different places within the paper, the calculations 
presented here serve as a proof of principle.}
\end{thebibliography}
\end{document}